\documentclass[a4paper, 11pt]{article}

\usepackage[utf8]{inputenc}
\usepackage{csquotes}
\usepackage[english]{babel}
\usepackage{expl3}
\usepackage{array}
\usepackage{url}
\usepackage{amssymb}
\usepackage{amsthm}
\usepackage{amsmath}
\usepackage{fullpage}
\usepackage{algorithm}
\usepackage{algpseudocode}
\usepackage{tabularx}
\usepackage{booktabs}
\usepackage{lipsum}
\usepackage{bm}
\usepackage{tikz}
\usetikzlibrary{arrows.meta, positioning}
\usepackage{siunitx}
\usepackage{graphicx}
\usepackage{xcolor}
\usepackage{caption}
\usepackage{subcaption}
\usepackage[section]{placeins}
\usepackage{authblk}
\usepackage{listings}

\definecolor{cobalt}{rgb}{0.0, 0.28, 0.67}

\newtheorem{thm}{Theorem}
\newtheorem{lem}[thm]{Lemma}

\theoremstyle{definition}

\DeclareUnicodeCharacter{2113}{\ensuremath{\ell}}

\renewcommand{\vec}[1]{\boldsymbol{#1}}

\makeatletter
\newcommand{\algorithmicbreak}{\textbf{break}}
\newcommand{\Break}{\State \algorithmicbreak}

\newcounter{phase}[algorithm]
\newlength{\phaserulewidth}
\newcommand{\setphaserulewidth}{\setlength{\phaserulewidth}}
\newcommand{\phase}[1]{%
  \vspace{1ex}
  \Statex\strut\refstepcounter{phase}\textbf{Phase~\thephase~--~#1}
  }
\makeatother
\setphaserulewidth{.7pt}

\title{Physics aware machine learning for micromagnetic energy minimization: recent algorithmic developments}
\author[a,b]{Sebastian Schaffer \thanks{\texttt{sebastian.schaffer@univie.ac.at}}} 
\author[a,d,e]{Thomas Schrefl}
\author[e]{Harald Oezelt}
\author[a,b,c]{Norbert J. Mauser}
\author[a,b]{Lukas Exl \thanks{\texttt{lukas.exl@univie.ac.at}}} 

\affil[a]{Research Platform MMM Mathematics - Magnetism - Materials, University of Vienna, Vienna, Austria}
\affil[b]{Wolfgang Pauli Institute, Vienna, Austria}
\affil[c]{Faculty of Mathematics, University of Vienna, Vienna, Austria}
\affil[d]{Christian Doppler Laboratory for magnet design through physics-informed machine learning, University for Continuing Education Krems, Wr. Neustadt, Austria}
\affil[e]{Department of Integrated Sensor Systems, University for Continuing Education Krems, Wr. Neustadt, Austria} 

\begin{document}
\maketitle
\noindent\textbf{Abstract.}
\noindent In this work, we explore advanced machine learning techniques for minimizing Gibbs free energy in full 3D micromagnetic simulations. Building on Brown’s bounds for magnetostatic self-energy, we revisit their application in the context of variational formulations of the transmission problems for the scalar and vector potential. To overcome the computational challenges posed by whole-space integrals, we reformulate these bounds on a finite domain, making the method more efficient and scalable for numerical simulation. Our approach utilizes an alternating optimization scheme for joint minimization of Brown’s energy bounds and the Gibbs free energy. The Cayley transform is employed to rigorously enforce the unit norm constraint, while $R$-functions are used to impose essential boundary conditions in the computation of magnetostatic fields.
Our results highlight the potential of mesh-free Physics-Informed Neural Networks (PINNs) and Extreme Learning Machines (ELMs) when integrated with hard constraints, providing highly accurate approximations. These methods exhibit competitive performance compared to traditional numerical approaches, showing significant promise in computing magnetostatic fields and the application for energy minimization, such as the computation of hysteresis curves. This work opens the path for future directions of research on more complex geometries, such as grain structure models, and the application to large scale problem settings which are intractable with traditional numerical methods.

\noindent\textbf{Keywords.}
micromagnetic energy minimization, magnetostatic energy bound, physics-informed neural network, boundary integral operator, extreme learning machine, vector potential\\

\noindent\textbf{Mathematics Subject Classification.} 	62P35,\, 68T07,\,  65Z05

\newpage

\section{Introduction}
Magnetostatics, a fundamental area of study in micromagnetics, focuses on understanding the behavior of magnetic fields and magnetization within materials. The minimization of magnetic energy, particularly Gibbs free energy, plays a critical role in determining stable magnetic configurations. Traditional numerical methods, such as the finite element method (FEM) \cite{schrefl2007numerical}  and finite difference methods (FDM) \cite{miltat2007numerical}, have long been used to solve magnetostatic problems, but these methods face challenges in efficiency and scalability, especially in three-dimensional domains with complex geometries and boundary conditions. In recent years, low-rank models were developed as a way to compress the high dimensional representation of the solution and perform computations in a reduced setting. Techniques such as multilinear low-rank tensor methods \cite{exl2014tensor} and spectral decomposition for computing the eigenmode evolution of the effective field operator \cite{d2009spectral, perna2022computational} exemplify these approaches. Kernel methods offer a way to compress magnetization data and capture magnetization dynamics within a low-dimensional latent space, enabling faster prediction of the system's response to changes in the external magnetic field \cite{exl2020learning, exl2021prediction, schaffer2021machine}. However, the supervised approach requires large amount of simulation data, which can be prohibitive.

Recently, there has been a growing interest in applying unsupervised Physics-Informed Neural Networks (PINNs) \cite{raissi2019physics} models to micromagnetic simulations, which provides promising alternatives to traditional numerical approaches and does not rely on precomputed data. PINNs have gained popularity for solving partial differential equations (PDEs) by embedding physical laws directly into the learning framework. Unlike conventional methods, they offer flexibility in the training process and can incorporate conditional parameters to learn a whole family of solutions \cite{kovacs2022conditional}. This can be particularly useful for inverse modelling. A related method, Extreme Learning Machines (ELMs) \cite{huang2011extreme}, allows for rapid training since only output parameters need to be optimized, offering computational efficiency while maintaining an accurate solution with a predetermined solution operator.
Hard-constrained PINNs and ELMs, incorporating R-functions \cite{rvachev1982theory, sukumar2022exact}, extend these methods by exactly enforcing boundary conditions within the model architecture. This avoids the need for soft constraint formulations and penalty terms. These approaches are particularly valuable in magnetostatics, where the enforcement of boundary conditions is crucial to accurate simulations. We apply these methods for the computational expensive computation of magnetostatic fields. 

One of the most challenging tasks in micromagnetic simulations is the minimization of the Gibbs free energy in full 3D systems. The main challenges lie in the costly computation of the magnetostatic fields and the enforcement of the unit norm constraint, which poses a difficult nonlinear optimization problem. 
Brown’s energy bounds \cite{brown1962magnetostatic}  offer theoretical limits for magnetostatic self-energy. However, in their plain form, they require the evaluation of whole space integrals, which is a challenging task for numerical simulations. Truncation approaches can be used, but they introduce an additional truncation error. Using the splitting ansatz proposed by Garcia-Cervera and Roma, we express these whole space integrals  solely on the finite domain of the magnetic body. With this description, we can use an alternating scheme for the joint optimization of the energy bounds and magnetization without considering the exterior space.
To account for the unit norm constraint, penalty frameworks \cite{schaffer2023physics} and a hard constraint ansatz using the Cayley transform \cite{schaffer2024constraint} have been proposed. Modern automatic differentiation (AD) frameworks allow the efficient computation of required differential operators and optimization of resulting objectives.

This article aims to explore the application of these modern machine learning methods, including PINNs and ELMs, in micromagnetic simulations, with a focus on the minimization of Gibbs free energy. By integrating traditional magnetostatic theory with state-of-the-art machine learning techniques, we aim to present new pathways for more efficient and scalable simulations.

\section{Physics-Informed Neural Networks}\label{sec:pinns}
Physics-informed machine learning and in particular a PINN is a novel approach to numerically solve a number of physical problems. 
Contrary to the supervised learning task, in a physics-informed setting, there is only partial or no label information available and the objective is designed to minimize the residual of some governing algebraic or differential equation. 

Due to the flexibility of neural networks, one can handle forward as well as inverse problems and integrate additional experimental or simulation data. PINNs have diverse applications in scientific and engineering domains such as fluid dynamics \cite{cai2021physics, raissi2020hidden}, material science \cite{zhang2022analyses}, geophysics and climate modelling \cite{kashinath2021physics}.

For instance, consider a Poisson equation with Dirichlet boundary conditions on the domain $\Omega$ with boundary $\partial\Omega$
\begin{equation}\label{eq:poisson}
    \begin{aligned}
        -\Delta u &= f \quad\, \text{in } \,\Omega \\
        u &= g \quad\, \text{on } \,\partial\Omega.
    \end{aligned}
\end{equation}
One possibility to find an approximate solution to this problem is to use a neural network ansatz $u_{\vec\omega} = u(\,\cdot\,;\vec\omega)$ which is parameterized by $\vec\omega$ and minimize the physics-informed objective
\begin{equation}\label{eq:poisson_pinn_loss}
    \mathcal{L}_{\text{PINN}}(\vec\omega) = \underbrace{\int_{\Omega} | \Delta u_{\vec\omega}(\vec x) + f(\vec x)|^2 \mathrm{d}\vec x}_{\mathcal{L}_{\text{PDE}}(\vec\omega)} + \lambda \underbrace{\int_{\partial \Omega} |u_{\vec\omega}(\vec x) - g(\vec x)|^2  \mathrm{d}s(\vec x)}_{\mathcal{L}_b(\vec\omega)},
\end{equation}
where $\lambda$ is a penalty parameter to account for boundary conditions. 

Partial derivatives can be computed by AD either in forward or reverse mode. If the input is low-dimensional, forward-mode AD can be used efficiently. Hence, the gradient $\nabla_{\vec \omega}\mathcal{L}_{\text{PINN}}$ can be computed with forward AD followed by reverse mode, i.e., \textit{reverse-over-forward}.

Choosing the penalty parameter $\lambda$ in \eqref{eq:poisson_pinn_loss} is not straightforward. One approach is to employ the squared penalty method, solving a series of optimization problems while gradually increasing the penalty parameter. However, while this may improve accuracy, it also significantly increases computational cost. Moreover, as the penalty parameter grows, the condition number of the Hessian worsens, leading to an ill-conditioned problem. Alternatively, instead of relying on the penalty formulation in \eqref{eq:poisson_pinn_loss}, one could adopt an augmented Lagrangian approach to minimize the PDE residual \cite{lu2021physics}.

A priori and a posteriori error estimates of residual based minimization of linear PDEs, using PINNs with soft constraints, have been derived in \cite{shin2023error}. This shows that a minimizer of \eqref{eq:poisson_pinn_loss} is actually a physically meaningful approximation to the true solution of the PDE.

Another example of PINNs would be the minimization of some energy functional with possible additional constraints in a penalty term, i.e.,
\begin{align}\label{eq:pinn-energy}
E(\vec \omega) = \int_\Omega \mathcal{E}\big(u_{\vec{\omega}}(\vec{x})\big) \mathrm{d}\vec x + \textrm{Penalty},
\end{align}
where $\mathcal E$ is the energy density.

To compute the respective integrals of the objective function accurately, one could for instance choose some quadrature rule. For arbitrary geometries it is usually easier to use Monte Carlo integration and select a set of collocation points $\mathcal{S} = \{\vec x_i\,|\, i=1, \dots, N\} \subseteq \Omega$, drawn from some distribution, to approximate the integral. A comprehensive study of different sampling strategies, as well as residual based sampling algorithms, is given in \cite{wu2023comprehensive} for various problem settings. It can be seen that quasi-random low discrepancy sequence leads to a lower relative error and importance sampling \cite{nabian2021efficient} further improves the performance and convergence of PINNs. However, this study does not take into account hard constraint PINNs (see section \ref{sec:hard_constraints}), but it is expected that these results also hold.
For an energy functional as \eqref{eq:pinn-energy}, where the energy density is non-zero, and the sampling distribution is non-uniform, importance sampling can be applied.

The main challenge is the minimization of the objective function. This is usually done with some variant of stochastic gradient descent, as well as second-order optimization methods like \textit{L-BFGS} or \textit{Trust-Region}. Further, hyperparameter optimization is a very challenging task for PINNs. It is not clear how to choose the perfect architecture to match the task at hand, and there is very little research regarding this issue \cite{sharma2023hyperparameter}. Overparametrization of the model is usually required to gain accuracy, resulting in a more complicated optimization problem which becomes more ill-conditioned with a growing number of parameters.

\subsection{Hard constraint formulation} \label{sec:hard_constraints}
An alternative approach to the soft constraint formulation, where the integrity of the solution is embedded in the governing objective, is the hard constraint formulation. Here, boundary conditions are strictly embedded into the architecture of the model. This transforms the optimization problem into an unconstrained one. For instance, in the case of problem \eqref{eq:poisson}, the model can be defined as.
\begin{equation}\label{eq:solution_structure_dirichlet}
    u_{\vec\omega}(\vec x) = \ell(\vec x) \hat u_{\vec\omega}(\vec x) + \tilde g(\vec x),
\end{equation}
where 
\begin{equation}
    \ell(\vec x) = \left\{
\begin{array}{ll}
      0 \quad\, \text{if}\,\, \vec x \in \partial\Omega \\
      > 0 \quad\, \text{if}\,\, \vec x \in \Omega.
\end{array} 
\right.
\end{equation}
Here, $\hat u_{\vec\omega}$ represents a neural network, while the function $\ell$ serves as an indicator. There are various ways to construct $\ell$, but as highlighted in \cite{sukumar2022exact}, it is advantageous if $\ell$ closely approximates the exact distance function near the boundary. The function $\tilde g$ is designed to interpolate $g$ at the boundary and to remain well-defined throughout the domain. It can be computed by means of transfinite interpolation \cite{rvachev2001transfinite, dyken2009transfinite} or trained with a neural network \cite{sheng2021pfnn}. Instead of minimizing the full loss function as in \eqref{eq:poisson_pinn_loss}, we focus solely on minimizing $\mathcal{L}_{\text{PDE}}$, since $\mathcal{L}_b$ is zero.

\section{Physics-Informed Extreme Learning Machines}\label{sec:elm}
An Extreme Learning Machine (ELM) is a type of neural network that contains only a single hidden layer. The input data is randomly mapped onto a latent space, meaning that only the output layer requires training. Formally, for an input $\vec{x} \in \mathbb{R}^d$, an ELM with $M$ hidden nodes and a scalar output can be described as
\begin{equation}
    u_{\vec\beta}(\vec{x}) = \sum_{i=1}^M \beta_i \sigma_i(\vec{x}) = \vec{\beta}^T \vec{z}(\vec{x}),
\end{equation}
where $\vec{z}$ represents the latent embedding of the hidden layer. For instance, $\sigma_i$ can be given by
\begin{itemize}
    \item an affine map followed by a nonlinear activation $\sigma$, i.e. $\sigma_i(\vec x) = \sigma(\vec w_i^T \vec x + b_i)$, where the input weights $(\vec w_i,b_i)$ are drawn from some distribution,
    \item a radial basis function, centered around a point $\vec w_i \in \Omega$, i.e. $\sigma_i(\vec x) = \exp(-\gamma\|\vec x - \vec w_i\|^2)$ with kernel parameter $\gamma$,
    \item or a basis of a B-spline function, leading to a sparse linear regression problem.
\end{itemize}

If the mean squared error (MSE) is to be minimized for some training set $\mathcal S = \{(\vec x_i, y_i)\,|\, i=1, \dots, N\} \subseteq \mathbb{R}^d \times \mathbb{R}$,
\begin{equation}\label{eq:mse-elm}
    \mathrm{MSE}(\vec \beta) = \frac{1}{2N} \sum_{i=1}^N (\vec\beta^T\vec z(\vec x_i) -  y_i)^2,
\end{equation}
the solution is given by the least squares problem
\begin{equation}
    H \vec \beta = \vec y,
\end{equation}
with
\begin{equation*}
    \begin{aligned}[c]
        H_{i,j} &= \sigma_j(\vec x_i) \\
        \vec y_i     &= y_i
    \end{aligned}
    \qquad\text{ for }\; i=1,\dots,N \text{ and } j=1,\dots,M.
\end{equation*}

with the solution operator being the pseudoinverse $H^\dagger=(H^T H)^{-1} H^T$.
If the size of the system becomes restrictive, one can use an incremental method to train the ELM \cite{huang2006universal}.

For an overparameterized model, computing the pseudoinverse can quickly become an ill-conditioned problem. Adding a Ridge regression term $\mu \|\vec\beta\|^2$ to \eqref{eq:mse-elm} can be very helpful in these cases. The simplest way is by modifying the pseudoinverse, i.e., when computing the Singular Value Decomposition (SVD) $H = USV^T$, with singular values $\sigma_1 \geq \dots \geq \sigma_M$, the regularized  solution operator is given by 
\begin{equation}
    H^\dagger_{\mu} = V S_\mu^{-1} U^T,
\end{equation}
where $S_\mu^{-1}$ is diagonal with the $i$\textsuperscript{th} diagonal entry given by $S_{\mu,ii}^{-1} = \sigma_i / (\sigma_i ^ 2 + \mu)$.

\subsection{Hard constraint ELMs}
Another possibility is the usage of a hard-constraint ELM ansatz, as in \eqref{eq:solution_structure_dirichlet}. The indicator function $\ell$ only acts on the activation functions and the ansatz becomes
\begin{equation}\label{eq:hard-constraint-elm-ansatz}
    u_{\vec\beta}(\vec x) = \ell(\vec x) \hat u_{\vec\beta}(\vec x) + \tilde g(\vec x) = \vec\beta^T\ell(\vec x)\vec z(\vec x) + \tilde g(\vec x).
\end{equation}
For a Poisson equation, as in \eqref{eq:poisson}, we have
\begin{equation}
    -\Delta(u_{\vec \beta}(\vec x) + \tilde g(\vec x)) = -\vec\beta^T\Delta\left(\ell(\vec x)\vec z(\vec x)\right) - \Delta \tilde g(\vec x) =  f(\vec x).
\end{equation}
Minimization of the squared residual
\begin{equation}\label{elm:squared-residual}
    \mathcal{L}_{\text{ELM}}(\vec\beta) = \int_{\Omega} | \vec\beta^T\Delta\left(\ell(\vec x)\vec z(\vec x)\right) + \Delta \tilde g(\vec x) + f(\vec x)|^2  \mathrm{d}\vec x,
\end{equation}
is equivalent to the solution of the system
\begin{equation}\label{eq:hard-constraint-elm-solution}
    H \vec \beta = \vec b,
\end{equation}
with
\begin{equation*}
\begin{aligned}[c]
    H_{ij} &= -\int_{\Omega}\Delta(\ell(\vec x)\sigma_i(\vec x))\Delta(\ell(\vec x)\sigma_j(\vec x))\mathrm{d}\vec x, \\
    \vec b_{i} &= \int_{\Omega}f(\vec x)\Delta(\ell(\vec x)\sigma_i(\vec x)) + \Delta \tilde g(\vec x) \Delta(\ell(\vec x)\sigma_i(\vec x))\mathrm{d}\vec x
\end{aligned}    
\begin{aligned}[c]
\quad&\text{ for }\; i=1,\dots,M \\ &\text{ and } j=1,\dots,M.    
\end{aligned}
\end{equation*}

Proper numerical integration of all those integrals might be very costly. A Monte-Carlo integration of \eqref{elm:squared-residual} is often easier with comparative performance. With the dataset of collocation points $\mathcal{S} = \{\vec x_i\,|\,i = 1, \dots, N\} \subseteq \Omega$ we get the objective
\begin{equation}\label{eq:elm-loss}
    \mathcal{L}_{\text{ELM}}(\vec\beta) = \frac{1}{N}\sum_{i=1}^N | \vec\beta^T\Delta\left(\ell(\vec x_i)\vec z(\vec x_i)\right) + \Delta \tilde g(\vec x_i) + f(\vec x_i)|^2.
\end{equation}
The minimizer of \eqref{eq:elm-loss} is given by the solution to the least squares problem
\begin{equation}\label{eq:hard-constraint-elm-solution-erm}
    A \vec \beta = \vec b,
\end{equation}
with
\begin{equation*}
    \begin{aligned}[c]
        A_{ij} &= -\Delta\left(\ell(\vec x_i)\sigma_j(\vec x_i)\right), \\
        \vec b_i &= f(\vec x_i) + \Delta \tilde g(\vec x_i) 
    \end{aligned}
    \qquad\text{ for }\; i=1,\dots,N \text{ and } j=1,\dots,M.
\end{equation*}
If this system is not too large, the pseudoinverse $A^\dagger$ defines the solution operator and can be precomputed using SVD. The best parameter $\vec\beta$ for some right-hand side $f$ is then given by a simple matrix multiplication. If the problem is ill conditioned, a Ridge regression term can be added. Further, for good performance and depending on the hidden layer activation, the input might be rescaled to $[-1, 1]^d$.

\subsubsection{Kernel perspective}
The regularized solution operator can also be computed with the dual formulation, i.e.,
\begin{equation}
    A^\dagger = (A^T A + \mu I)^{-1} A^T = A^T (A A^T + \mu I)^{-1},
\end{equation}
by application of the push-through identity. This is equivalent to Kernel Ridge Regression (KRR) \cite{murphy2012machine} with the kernel function given by
\begin{equation}
    k(\vec x_1, \vec x_2) = \Delta\left(\ell(\vec x_1)\vec z(\vec x_1)\right)^T \Delta\left(\ell(\vec x_2)\vec z(\vec x_2)\right).
\end{equation}
Thus, the kernel matrix is given by $K=AA^T$. 

There is a clear connection of ELMs and the low-rank KRR \cite{schaffer2021machine, exl2021prediction, williams2000using}. Consider a KRR with kernel function $a(\vec x_i,\vec x_j) = \ell(\vec x_i)\psi(\vec x_i)\cdot\ell(\vec x_j)\psi(\vec x_j)$, where $\psi:\Omega\rightarrow\mathcal{F}_{\Omega}, \vec x\mapsto \psi(\vec x)$ is a feature map which transforms the input to a reproducing kernel Hilbert space (RKHS) and $\cdot$ denotes the inner product in this space.
The regularized physics-informed loss is given by 
\begin{equation}\label{eq:krr-loss}
    \mathcal{L}_{\text{KRR}}(\vec\xi) = \frac{1}{N}\sum_{i=1}^N | \Delta\left(\ell(\vec x_i)  \psi(\vec x_i)\right)\cdot\vec\xi + \Delta \tilde g(\vec x_i) + f(\vec x_i)|^2 + \mu(\vec\xi\cdot\vec\xi).
\end{equation}
Let $\psi_\ell(\vec x) = \Delta\left(\ell(\vec x)  \psi(\vec x)\right)$ and denote $\Psi[\mathcal{S}]$ as the matrix of size $N \times \mathrm{dim}(\mathcal{F}_{\Omega})$ which holds the samples mapped to the RKHS
\begin{equation}
    \Psi[\mathcal{S}] := [\psi_\ell(\vec x_1)|\dots|\psi_\ell(\vec x_N)]^T.
\end{equation}
The dual solution would then be given by
\begin{equation}
    \vec\xi = \Psi[\mathcal{S}]^T (\Psi[\mathcal{S}] \cdot \Psi[\mathcal{S}] + \mu I)^{-1} \vec b = \Psi[\mathcal{S}]^T (G[\mathcal{S},\mathcal{S}] + \mu I)^{-1} \vec b,
\end{equation}
where $G[\mathcal{S},\mathcal{S}]$ is the corresponding kernel matrix. If the kernel matrix $G[\mathcal{S},\mathcal{S}]$ is now approximated with a low-rank version of rank $M$, i.e,
$G[\mathcal{S},\mathcal{S}] \approx \widetilde \Psi[\mathcal{S}] \widetilde \Psi[\mathcal{S}]^T$ with $\widetilde \Psi[\mathcal{S}]$ being a low-rank matrix of size $N\times M$ of $\Psi$, the solution operator has the same form as for the ELM ansatz.
The problem with this ansatz is the computation of the Laplace operator, since it is not clear what the corresponding kernel to $\Delta\left(\ell(\vec x_i)\psi(\vec x_i)\right) \Delta\left(\ell(\vec x_j)  \psi(\vec x_j)\right)$ should be. The ELM solves this by mapping the input directly onto a randomized low-rank RKHS. The feature space mapping $\widetilde\Psi[\mathcal{S}]$ is directly given by the ELM's
hidden layer output. 

In our experiments, selecting for instance a radial basis activation function with $\ell(\vec x)\sigma_i(\vec x) = \ell(\vec x)\exp(-\gamma \|\vec x - \vec w_i\|^2)$ with $\vec w_i \in  \Omega$ being the weight of the $i$\textsuperscript{th} node of the ELM, achieves similar performance to other activation functions, such as \texttt{tanh} or \texttt{gelu} with the advantage that the weights can be drawn from the domain and there is no intercept term required. For difficult functions, such as the approximation of a vortex state with vector potential (see Section~\ref{sec:vortex-state}), this kind of ELM even outperforms ELMs with other activations. The disadvantage lies in the selection of the kernel parameter $\gamma$ which requires hyperparameter tuning.

\subsubsection{Neural operator perspective}
The physics-informed ELM can also be seen as a kind of operator network, like Deep Operator Networks {DeepONet} \cite{lu2019deeponet}. A DeepONet $G$ has the basic structure
\begin{equation}
    G(\vec x;\,u) = B(u)\cdot T(\vec x),
\end{equation}
where $B$ is the so-called branch network and $T$ the trunk network. The branch network takes as input a function $u$ (or discrete measurements of the function) and the trunk network takes a point from the input domain. The result is given by the inner product of the branch and the trunk. The DeepONet can the then be trained in a physics-informed setting for various input functions $u$, sampled from some function space for the goal to approximate the respective inverse operator. For an ELM and the objective \eqref{eq:elm-loss}, the branch is given by $B(f, g) = A^\dagger \vec b(f, g)$ and the trunk by $T(\vec x) = \ell(\vec x)\vec z(\vec x)$. The main difference is, that only a shallow network is applied and there is no complicated training procedure required for the determination of the solution operator. Pre-training a DeepONet might be a versatile alternative to ELMs. Especially, Graph Neural Networks \cite{wu2020comprehensive} are of interest for the task of learning the solution operator \cite{sharma2024graph, cho2024graphdeeponet}. The main challenge lies in the training of the model, which requires sampling from some function space. ELMs could be utilized for efficient sampling of such a space. Especially, training of the branch network can be challenging.
\newline

\noindent All in all, those seemingly simple ELM models are very powerful function approximators and are capable to solve quite complex problems with ease. For small scale problems, the solution can be computed without the need of the weak formulation, allowing a stable computation without preconditioning. However, for larger irregular domains, it is still questionable how to select the input weights properly. A Gradient Boosting technique could be applied to solve for consecutive small ELMs and the error information might be applied as information for the input weight distribution.

\section{Magnetostatics with PINNs}\label{sec:magnetostatics}

\noindent Magnetostatics focuses on the study of magnetic fields in systems with steady currents. The fundamental equations governing magnetostatics are derived from Maxwell's equations and describe how magnetic fields are generated by steady currents and magnetic materials. Solving these equations typically involves complex boundary conditions and interactions between magnetic domains, which has traditionally required numerical methods such as the finite difference (FDM) or finite element methods (FEM).

Recently, the advent of PINNs has introduced a novel approach to solving magnetostatic problems. PINNs can be used to model continuous magnetization configurations commonly used in micromagnetism \cite{exl2020micromagnetism}. This approach not only simplifies the computational process but also offers flexibility in incorporating additional conditional parameters such as exchange length. Moreover, as described in this chapter, they can efficiently handle the computationally intensive stray field problem by employing techniques like ELMs and splitting methods, further enhancing their applicability to magnetostatic simulations.

\subsection{The magnetostatic self-energy}
An important aspect of magnetostatics is the calculation of magnetostatic self-energy, which represents the energy stored in the magnetic field due to the magnetization of the material. This self-energy is a crucial factor in determining the stability and equilibrium configurations of magnetic systems, influencing how magnetic domains form and interact.

The magnetostatic self-energy or demagnetizing energy $E_d$ of a magnet $\Omega \subset \mathbb{R}^3$ is usually computed via the stray field $\vec H_d = M_s \vec h_d$ through \cite{brown1963micromagnetics}
\begin{align}\label{eq:demag-energy}
    E_d(\vec m) = -\frac{\mu_0 M_s^2}{2} \int_{\Omega} \vec m \cdot \vec h_d \, \mathrm{d}\vec x = \frac{\mu_0 M_d^2}{2} \, e_d(\vec m), 
\end{align}
where $\vec m \in L^2(\mathbb R^3)^3$ is the magnetization configuration, which is non-zero in the bounded domain $\Omega$ and zero in the exterior domain $\overline\Omega^c = \mathbb{R}^3 \setminus \overline{\Omega}$, $M_s$ the saturation magnetization, $\mu_0$ the vacuum permeability, $\vec h_d \in L^2(\mathbb R^3)^3$ the dimensionless stray field and $e_d$ the reduced self-energy in dimensions of volume. 

Accurate computation of the stray field is crucial for many numerical algorithms in micromagnetism. This step enables precise mathematical modeling of magnetic system behaviors, which is essential for designing highly efficient magnetic devices or permanent magnets \cite{suess2018topologically, huber20173d,fischbacher2018micromagnetics}.

The stray field $\vec h_d = -\nabla \phi_d$ is the solution of the whole space Poisson problem 
\begin{align}\label{eq:poisson-scalar-potential}
\Delta \phi_d = \nabla \cdot \vec m\quad \text{in }\mathbb{R}^3
\end{align}
for the scalar potential $\phi_d$. 
The Poisson equation \eqref{eq:poisson-scalar-potential} follows from the magnetostatic Maxwell's equations,
\begin{eqnarray}\label{eq:static-maxwell}
\begin{aligned}
  \vec b_d &\,= \vec m + \vec h_d\\  
  \nabla \cdot \vec b_d &\, = 0 \\
  \nabla \times \vec h_d &\, = \boldsymbol{0},
\end{aligned}
\end{eqnarray}
where $\vec b_d = \vec B_d / (M_s\mu_0)$ is the (dimensionless) magnetic induction (or magnetic flux density) $\vec B_d$. The fields $\vec h_d = \vec h_d(\vec m)$ and $\vec b_d=\vec b_d(\vec m)$ depend (linearly) on $\vec m$, but we omit this dependency for the sake of simpler notation.
The components $\vec b_d\cdot \vec n$ and $\vec h_d\times\vec n$, with $\vec n$ denoting the outward unit normal vector, are continuous across the surface of $\Omega$ \cite{brown1962magnetostatic}. 
The fields $\vec h_d$ and $\vec b_d$ form a Helmholtz decomposition of the magnetization $\vec m$ in whole space into a sum of an irrotational and solenoidal part \cite{brown1962magnetostatic}, which are $L^2$-orthogonal in whole space, that is,
\begin{align}\label{eq:bh-orthogonality}
    \int_{\mathbb{R}^3} \vec h_d\cdot \vec b_d\, \mathrm{d}\vec x = 0.
\end{align}
Using this orthogonality and the relations $\vec b_d = \vec m + \vec h_d$ in \eqref{eq:static-maxwell} one can express the demagnetizing energy \eqref{eq:demag-energy} in the equivalent forms
\begin{equation}\label{eq:demag-energies}
    \begin{aligned}
    e_d(\vec m) &=  -\int_{\Omega} \vec m \cdot \vec h_d \, \mathrm{d}\vec x = \int_{\mathbb{R}^3} \|\vec h_d\|^2 \, \mathrm{d}\vec x  \\
    & =  \int_{\Omega} \|\vec m\|^2  -  \vec m \cdot \vec b_d \, \mathrm{d}\vec x = \int_{\Omega} \|\vec m\|^2 \, \mathrm{d}\vec x - \int_{\mathbb{R}^3} \|\vec b_d\|^2 \, \mathrm{d}\vec x,
    \end{aligned}
\end{equation}
where integration involving $\vec m$ is over $\Omega$ only and $\int_\Omega \|\vec m\|^2\,dx = V$ the volume of $\Omega$. 
Opposed to $\vec h_d$ being the solution of \eqref{eq:poisson-scalar-potential}, the magnetic induction $\vec b_d = \nabla \times \vec A_d$ is the solution of the whole space vector Poisson equation
\begin{align}\label{eq:poisson-vec-potential}
\nabla \times (\nabla \times \vec A_d) = \nabla \times \vec m\quad \text{in }\mathbb{R}^3
\end{align}
for the vector potential $\vec A_d$. There is freedom of gauge for the vector potential, since one can add an arbitrary gradient field, i.e., $\vec b_d = \nabla \times (\vec A_d + \nabla \Psi)$. Usually the \textit{Coulomb gauge} 
\begin{align}
\nabla \cdot \vec A_d = 0
\end{align} 
is chosen, which reduces \eqref{eq:poisson-vec-potential} to
\begin{align}\label{eq:poisson-vec-potential2}
\Delta  \vec A_d = -\nabla \times \vec m\quad \text{in }\mathbb{R}^3.
\end{align}

Splitting the interior and the exterior part of the scalar potential, $\phi_d = \phi^{int}_d$ in $\Omega$ and $\phi_d = \phi^{ext}_d$ in $\overline\Omega^c$, \eqref{eq:poisson-scalar-potential} can be formulated as a transmission problem \cite{carstensen1995adaptive,exl2018magnetostatic}
\begin{equation}\label{eq:stray-field}
    \begin{aligned}
    -\Delta \phi_d &= -\nabla\cdot \vec m \qquad &&\text{in} \; \Omega \subset \mathbb{R}^3, \\
    -\Delta \phi_d &= 0 &&\text{in} \; \overline\Omega^c, \\
    [\phi_d] &= 0 &&\text{on} \; \partial\Omega, \\
    \left[D_{\vec n}\phi_d\right] &= - \vec m \cdot \vec n &&\text{on} \; \partial\Omega, \\
    \phi_d(\vec x) &= \mathcal{O}\left(\frac{1}{\|\vec x\|}\right) &&\text{as} \; \|\vec x\|\rightarrow \infty.
    \end{aligned}
\end{equation}
The jump conditions $[\,\cdot\,]$ describe the difference of the exterior and interior behavior at the boundary of the magnet, i.e.,
\begin{equation}\label{eq:jump-conditions}
\begin{aligned}\relax
    [\phi_d] &= \left.(\phi^{ext}_d - \phi_d^{int})\right|_{\partial\Omega} \\
    [D_{\vec n}\phi_d] &= \left.(D_{\vec n}\phi^{ext}_d - D_{\vec n}\phi^{int}_d)\right|_{\partial\Omega}.
\end{aligned}
\end{equation}
Further, the normal derivative is denoted by $D_{\vec n}$. For \eqref{eq:stray-field} to be well-posed, the boundary $\partial\Omega$ needs to be Lipschitz continuous. For grain structure modeling, this is usually true.
In the same way, the Poisson equation of the vector potential \eqref{eq:poisson-vec-potential2} gives \cite{exl2018magnetostatic}
\begin{equation}\label{eq:b-field}
    \begin{aligned}
    \Delta \vec A_d &= -\nabla\times \vec m \qquad &&\text{in } \Omega, \\
    \Delta \vec A_d &= \vec 0 &&\text{in }\overline\Omega^c, \\
    [\vec A_d] &= \vec 0 &&\text{on }\partial\Omega, \\
    \left[D_{\vec n}\vec A_d\right] &= - \vec m \times \vec n &&\text{on }\partial\Omega, \\
    (\vec A_d)_j(\vec x) &= \mathcal{O}\left(\frac{1}{\|\vec x\|}\right) 
 &&\text{as } \|\vec x\|\rightarrow \infty, \, j=1,2,3.
    \end{aligned}
\end{equation}

The transmission problems \eqref{eq:stray-field} or \eqref{eq:b-field} can be solved in different ways numerically. Finite element methods usually solve the weak formulation for certain splitting of the potentials \cite{fredkin1990hybrid, garcia2006adaptive, schrefl2007numerical, exl2014non}, finite difference methods approximate the integral representations of the solution \cite{miltat2007numerical}, 
\begin{equation}\label{eq:stray-field-solution}
    \phi_d(\vec x) = -\frac{1}{4\pi}\left(\int_\Omega \frac{\nabla\cdot \vec m(\vec y)}{\|\vec x - \vec y\|}\mathrm{d}\vec y
    - \int_{\partial\Omega} \frac{\vec m(\vec y)\cdot \vec n(\vec y)}{\|\vec x - \vec y\|}\mathrm{d}s(\vec y)\right).
\end{equation}
However, there are significant challenges in implementing this solution. Firstly, the computation of the convolution integral across the entire domain is highly resource-intensive. Secondly, obtaining the gradient of \eqref{eq:stray-field-solution} is typically done with numerical differentiation, which can lead to discretization and round-off errors.
Recent neural network methods take a mixed approach \cite{schaffer2023physics,schaffer2024constraint}. In whatever way the solution is obtained, the computation of the $\vec h_d$-field (or $\vec b_d$-field) is usually the most time-consuming part in micromagnetic modelling due to its non-local nature \cite{abert2013numerical}.

\subsection{Brown's energy bounds}
In the 60s, W.F. Brown suggested simple sharp bounds for the magnetostatic energy only in terms of field variables with no dependency on $\vec m$ and free from long-range interactions \cite{brown1962magnetostatic,brown1963micromagnetics}. For any irrotational field $\vec h \in L^2(\mathbb{R}^3)^3$ and solenoidal field $\vec b\in L^2(\mathbb{R}^3)^3$ holds 
\begin{align}\label{eq:bounds}
   \int_{\Omega} \|\boldsymbol{m}\|^2 \, \mathrm{d}\vec x - \int_{\mathbb{R}^3} \|\boldsymbol{h} + \boldsymbol{m}\|^2 \, \mathrm{d}\vec x  \, \leq \, e_d(\boldsymbol{m}) \, \leq \, \int_{\mathbb{R}^3} \|\boldsymbol{b} - \boldsymbol{m}\|^2 \, \mathrm{d}\vec x.
\end{align}
The upper bound is derived by adding  $\int_{\mathbb{R}^3} \|\boldsymbol{b}_d - \boldsymbol{b}\|^2 \, dx$ to the energy, while the lower bound results from subtracting $\int_{\mathbb{R}^3} \|\boldsymbol{h}_d - \boldsymbol{h}\|^2 \, dx$ from the energy \cite{asselin1986field} and using $\vec b_d = \vec m + \vec h_d$ as well as orthogonality of irrotational and solenoidal fields \eqref{eq:bh-orthogonality}.
In order to compute the energy as well as the respective stray field $\vec h_d$, one can  maximize the lower bound with respect to a scalar potential $\phi$ for the ansatz $\vec h = -\nabla \phi$. Similarly, minimizing the upper bound with $\vec{b} = \nabla \times \vec{A}$ with a vector potential $\vec A$ gives the respective magnetic induction $\vec b_d$ in the optimum.

While those energy bounds have been shown to be applicable in a physics-informed setting \cite{kovacs2022magnetostatics} for 2d, in practice however, the integration over whole space in the bounds \eqref{eq:bounds} causes problems. We will come back to this issue in Section~\ref{sec:brown-revisited}. For now, we will use these bounds to derive some basic results.

\subsubsection{Weak formulation}
A variational form of the transmission problem \eqref{eq:stray-field} involving the scalar potential is derived by utilizing partial integration and the jump condition $\left[D_{\vec n}{\phi_d}\right] = - \vec m \cdot \vec n $ at the boundary:
\begin{equation}\label{eq:weakpoisson-stray-field}
\begin{aligned}
    &\text{For $\vec m\in \big(H^1(\Omega)\big)^3$, find $\phi_d \in X_\phi$ such that for all $\phi \in X_\phi$ there holds}\\
    &\int_{\mathbb{R}^3} \nabla \phi_d \cdot \nabla \phi \, \mathrm{d}\vec x = \int_{\Omega} \vec m \cdot \nabla \phi\, \mathrm{d}\vec x,
\end{aligned}
\end{equation}
where $X_\phi$ is a suitable weighted Sobolev space \cite{sauter2011boundary} which incorporates the continuity across the boundary and the decay behavior.
There exists a unique solution of \eqref{eq:weakpoisson-stray-field} in $X_\phi$. Solving the weak form \eqref{eq:weakpoisson-stray-field} is equivalent to maximizing the following functional over $X_\phi$
\begin{align}\label{eq:phi-energy-func}
   e_\phi(\phi; \,\vec m) = -\int_{\mathbb{R}^3} \|\nabla \phi\|^2 \, \mathrm{d}\vec x + 2\int_{\Omega} \vec m \cdot \nabla \phi\, \mathrm{d}\vec x.
\end{align}

The variational formulation for the vector potential is derived similarly, using the jump condition $\left[D_{\vec n}{\vec A_d}\right] = - \vec m \times \vec n$:
\begin{equation}\label{eq:weakpoisson-vec-potential}
\begin{aligned}
    &\text{For $\vec m\in \big(H^1(\Omega)\big)^3$, find $\vec A_d \in X_{\vec A}$ such that for all $\vec A \in X_{\vec A}$ there holds}\\
    &\int_{\mathbb{R}^3} \nabla \vec A_d : \nabla \vec A \, \mathrm{d}\vec x = \int_{\Omega} \vec m \cdot (\nabla \times \vec A) \, \mathrm{d}\vec x,
\end{aligned}
\end{equation}
where $\nabla \vec A_d : \nabla \vec A=Tr[\nabla (\vec A_d)^T \nabla \vec A]$ denotes double contraction. Solving \eqref{eq:weakpoisson-vec-potential} is equivalent to minimizing
\begin{align}\label{eq:A-energy-func}
   e_{\vec A}(\vec A; \,\vec m) = \int_{\Omega} \|\vec m\|^2 \, \mathrm{d}\vec x + \int_{\mathbb{R}^3} \|\nabla \vec A\|^2 \, dx - 2 \int_{\Omega} \vec m \cdot (\nabla \times \vec A)\, \mathrm{d}\vec x
\end{align}
over $X_{\vec A}$, where $ \int_{\Omega} \|\vec m\|^2 \, \mathrm{d}\vec x$ is a constant and $\|\nabla \vec A\|^2 = \nabla \vec A : \nabla \vec A$.

The energy functionals \eqref{eq:phi-energy-func} and \eqref{eq:A-energy-func} correspond to the lower and upper bound of \eqref{eq:bounds}
\begin{equation}
    e_\phi(\phi; \,\vec m) \leq e_d(\vec m) \leq e_{\vec A}(\vec A; \,\vec m).
\end{equation}

In numerical computations, we do not actually solve the potential equations in the infinite dimensional spaces above, but rather assume a (finite dimensional) approximation space $\widetilde X_\phi \subset X_\phi$.
Hence, assuming an exact solution $\widetilde \phi_d \in \widetilde X_\phi$ with $\widetilde{\vec h}_d = -\nabla\widetilde \phi_d$ of the corresponding weak form \eqref{eq:weakpoisson-stray-field}, that is, we solve the (Galerkin) problem:
\begin{equation}\label{eq:weakpoisson-stray-field-galerkin}
\begin{aligned}
    &\text{For $\vec m\in \big(H^1(\Omega)\big)^3$, find $\widetilde\phi_d \in \widetilde X_\phi$ such that for all $\phi \in \widetilde X_\phi$ there holds}\\
    &\int_{\mathbb{R}^3} \nabla \widetilde\phi_d \cdot \nabla \phi \, \mathrm{d}\vec x = \int_{\Omega} \vec m \cdot \nabla \phi\, \mathrm{d}\vec x.
\end{aligned}
\end{equation}
Equivalently we can maximize the energy functional \eqref{eq:phi-energy-func} over $\widetilde X_\phi$,
\begin{equation}
    \widetilde \phi_d = \arg\max_{\phi\in\widetilde X_\phi} e_\phi(\phi;\, \vec m)
\end{equation}
Applying this solution to \eqref{eq:phi-energy-func} gives
\begin{align}\label{eq:approx-stray-field-energy}
   e_\phi(\widetilde \phi_d; \,\vec m) = -\int_{\Omega} \vec m \cdot \widetilde{\vec h}_d\, \mathrm{d}\vec x.
\end{align}
Therefore, by using the $L^2$-orthogonality \eqref{eq:bh-orthogonality} and the first relation in \eqref{eq:static-maxwell}
\begin{equation}
    |e_d(\vec m) - e_\phi(\widetilde \phi_d; \,\vec m)| = |\int_{\Omega} \vec m \cdot (\vec h_d - \widetilde{\vec h}_d)\, \mathrm{d}\vec x| = |\int_{\mathbb{R}^3} \vec h_d  \cdot (\vec h_d - \widetilde{\vec h}_d)\, \mathrm{d}\vec x|
\end{equation}
and the Cauchy-Schwarz inequality, the error of the self energy can be bound by
\begin{equation}\label{eq:error-energy}
    \frac{|e_d(\vec m) - e_\phi(\widetilde \phi_d; \,\vec m)|}{\sqrt{e_d(\vec m)}} \leq 
    \left(\int_{\mathbb{R}^3} \|\vec h_d - \widetilde{\vec h}_d\|^2\, \mathrm{d}\vec x\right)^{\frac{1}{2}}.
\end{equation}

With the same line of reasoning, for an exact solution $\widetilde{\vec A}_d$ with $\widetilde{\vec b}_d = \nabla \times \widetilde{\vec A}_d$ of \eqref{eq:weakpoisson-vec-potential} on a finite dimensional subspace $\widetilde X_{\vec A} \subset X_{\vec A}$, one finds
\begin{equation}
    \frac{|e_d(\vec m) - e_{\vec A}(\widetilde{\vec A}_d; \,\vec m)|}{\sqrt{\int_{\Omega} |\vec m|^2 \, \mathrm{d}\vec x - e_d(\vec m)}} \leq 
    \left(\int_{\mathbb{R}^3} \|\vec b_d - \widetilde{\vec b}_d\|^2\, \mathrm{d}\vec x\right)^{\frac{1}{2}}.
\end{equation}
Therefore, the deviation of the energy values can be estimated by the $L^2$-error resulting from the respective field computation.

Further, since the exact solution on an approximate solution space of the weak form is in the infinite-dimensional spaces $X_\phi$ and $X_{\vec A}$ respectively, the self-energy for the scalar potential is underestimated while the energy for the vector potential is overestimated, i.e.,
\begin{equation}\label{eq:energy-bounds-weal-sol}
     -\int_{\Omega} \vec m \cdot \widetilde{\vec h}_d\, \mathrm{d}\vec x \leq e_\phi(\phi_d; \,\vec m) = e_d(\vec m) = e_{\vec A}(\vec A_d; \,\vec m) \leq  \int_{\Omega} \|\vec m\|^2 - \vec m \cdot \widetilde{\vec b}_d\, \mathrm{d}\vec x.
\end{equation}
However, our assumption of an exact solution of the approximate formulation is not realistic in numerical computations in general, where the whole space problem \eqref{eq:weakpoisson-stray-field-galerkin} will hardly hold exactly, and likewise \eqref{eq:approx-stray-field-energy}. Hence, \eqref{eq:energy-bounds-weal-sol} might be violated from both sides. 
Nevertheless, in this case where we assume only an approximation $\widetilde{\phi} \approx \widetilde{\phi}_d$ and $\widetilde{\vec A} \approx \widetilde{\vec A}_d$, the lower and upper bounds are still guaranteed to hold,
\begin{equation}\label{eq:energy-bounds-with-func}
    e_\phi(\widetilde \phi; \,\vec m) \leq  e_d(\vec m)  \leq e_{\vec A}(\widetilde{\vec A}; \,\vec m).
\end{equation}


For the evaluation of the self-energy, it is necessary to compute the stray field or the magnetic induction within the magnetic domain. However, this requires a solution to the whole space problem \eqref{eq:stray-field}. One could for instance use a truncation approach and minimize the upper bound in \eqref{eq:bounds}. In \cite{kovacs2022magnetostatics} this was done in 2d. In 3d, however, it turned out to be much more challenging, and we were not able to verify this approach. This could be due to the jump conditions which are not automatically satisfied. Especially the jump in the normal derivative can not be satisfied with a simple neural network ansatz. An easier and more flexible approach is described in the following section and in section~\ref{sec:opt_bounds}.

\subsection{Splitting ansatz of Garcia-Cervera and Roma}\label{sec:garcia-cervera-roma}
To find a solution to the whole space transmission problem \eqref{eq:stray-field}, it can be expressed as two easier sub-problems which only require the solution within the magnetic domain. Therefore, the splitting ansatz of Garcia-Cervera and Roma is employed. It can be used for the scalar and the vector potential alike. For the scalar potential, it is split into two parts $\phi = \phi_1 + \phi_2$ with
\[
\phi_1 = \bigg\{
\begin{array}{ll}
\phi_1^{int} & \text{in}\; \Omega\\ 
\phi_1^{ext} & \text{in}\; \overline\Omega^c\\ 
\end{array},\; \phi_2 = \bigg\{
\begin{array}{ll}
\phi_2^{int} & \text{in}\; \Omega\\ 
\phi_2^{ext} & \text{in}\; \overline\Omega^c\\ 
\end{array},
\]
and $\phi_1^{int},\; \phi_2^{int}\in  H^1(\Omega)$ and $\phi_1^{ext},\;\phi_2^{ext} \in H^1_{loc}(\overline\Omega^c)$. 

The component $\phi_1$ satisfies the homogeneous Dirichlet Poisson problem, i.e.
\begin{equation}\label{eq:phi1-poisson}
    \begin{aligned}
        -\Delta \phi_1^{int} &= - \nabla \cdot \vec m &\text{in}\; \Omega, \\ 
                \phi_1^{int} &= 0 &\text{on}\; \partial \Omega \\
                \phi_1^{ext} &= 0 &\text{in}\; \overline\Omega^c,
    \end{aligned}
\end{equation}
whereas the second component $\phi_1$ fulfills the Laplace equation
\begin{equation}\label{eq:phi2-interface}
    \begin{aligned}
        -\Delta \phi_2^{int} &= 0 &\text{in}\; \Omega, \\ 
        -\Delta \phi_2^{ext} &= 0 &\text{in}\; \overline\Omega^c, \\ 
        [\phi_2] &= 0 &\text{on}\; \partial \Omega, \\
        \left[D_{\vec n}\phi_2\right] &= -\vec m \cdot \vec n + D_{\vec n}\phi_1^{int} &\text{on}\; \partial \Omega, \\
        \phi_2^{ext}(\vec x) &= \mathcal{O}\left(\frac{1}{\|\vec x\|}\right) &\text{as} \; \|\vec x\|\rightarrow \infty.
    \end{aligned}
\end{equation}
It is easy to verify that a solution to \eqref{eq:phi1-poisson} and \eqref{eq:phi2-interface} satisfies the jump conditions in \eqref{eq:stray-field} since $[\phi_1] = 0$ and $[D_{\vec n}\phi_1] = - D_{\vec n}\phi_1^{int}$.

The solution of \eqref{eq:phi2-interface} can be expressed with the \textit{single layer potential}
\begin{equation}\label{eq:single_layer_pot}
    \phi_2^{ext}(\vec x) = \phi_\nu(\vec x; \phi_1,\vec m) := \frac{1}{4\pi}\int_{\partial\Omega} \frac{(\vec m\cdot \vec n - D_{\vec n}\phi_1^{int})(\vec y)}{\|\vec x - \vec y \|}\, \mathrm{d}s(\vec y).
\end{equation}

Likewise, the vector potential can be split in $\vec A = \vec A_1 + \vec A_2$ with
\begin{equation}\label{eq:A1-poisson}
    \begin{aligned}
        \Delta \vec A_1^{int} &= - \nabla \times \vec m &\text{in}\; \Omega, \\ 
                \vec A_1^{int} &= 0 &\text{on}\; \partial \Omega \\
                \vec A_1^{ext} &= 0 &\text{in}\; \overline\Omega^c,
    \end{aligned}
\end{equation}
and 
\begin{equation}\label{eq:A2-interface}
    \begin{aligned}
        \Delta \vec A_2^{int} &= 0 &\text{in}\; \Omega, \\ 
        \Delta \vec A_2^{ext} &= 0 &\text{in}\; \overline\Omega^c, \\ 
        [\vec A_2] &= 0 &\text{on}\; \partial \Omega, \\
        \left[D_{\vec n}\vec A_2\right] &= -\vec m \times \vec n + D_{\vec n}\vec A_1^{int} &\text{on}\; \partial \Omega, \\
        (\vec A_2^{ext})_j(\vec x) &= \mathcal{O}\left(\frac{1}{\|\vec x\|}\right) ,\; j=1,2,3, &\text{as} \; \|\vec x\|\rightarrow \infty.
    \end{aligned}
\end{equation}
The solution of $\vec A_2^{ext}$ is then given by 
\begin{equation}\label{eq:A2-single_layer_pot}
    \vec A_2^{ext}(\vec x) = \vec A_\nu(\vec x;\vec A_1^{int},\vec m) := \frac{1}{4\pi}\int_{\partial\Omega} \frac{(\vec m \times \vec n - D_{\vec n}\vec A_1^{int})(\vec y)}{\|\vec x - \vec y \|}\, \mathrm{d}s(\vec y).
\end{equation}

With the single layer potential, the solution of the exterior domain can be projected onto the boundary. Only the interior domain $\Omega$ needs to be considered and therefore, from now on, we dismiss the superscript $int$ and $ext$.

To compute a solution to the stray field, one could proceed as follows. First, compute a solution $\phi_1$ to \eqref{eq:phi1-poisson}. Second, use the retrieved solution to compute $\phi_2$ with \eqref{eq:single_layer_pot} and third, compute $\vec h_d=-\nabla (\phi_1 + \phi_2)$.
The computation of the stray field with a hard constrained ELM ansatz is summarized in Algorithm~\ref{alg:elm-stray-field}.
\begin{algorithm}
\caption{Computation of the stray field with a hard constrained ELM ansatz}
\label{alg:elm-stray-field}
\begin{algorithmic}
\Require Magnetization $\vec m$, ADF $\ell: \mathbb{R}^3 \rightarrow \mathbb{R}$, ELM embedding $\vec z: \Omega \rightarrow \mathbb{R}^M$, collocation points $\{\vec x_1, \dots, \vec x_N\} \subset \Omega$ and Ridge regularization parameter $\mu$
\phase{Precomputation:}
\State $A_{ij} = -\Delta\left(\ell(\vec x_i)\vec z(\vec x_i)_j\right)$ for $i=1,\dots,N$ and $j=1,\dots,M$
\State $USV^T = A$   \Comment{compute truncated SVD}
\State $A^\dagger = V S (S^2 + \mu I)^{-1} U^T$ \Comment{compute solution operator for $\phi_1$}
\phase{Computation of $\phi_1$:}
\State  $\vec b_i = -\nabla \cdot \vec m(\vec x_i)$ for $i=1,\dots,N$ \Comment{compute right hand side}
\State $\vec \beta = A^\dagger \vec b$
\State $\widetilde{\phi}_1(\,\cdot\,) = \vec \beta^T\ell(\,\cdot\,)\vec z(\,\cdot\,)$ \Comment{approximate solution to $\phi_1$}
\phase{Computation of $\phi_2$:}
\State $\widetilde{\phi}_2(\,\cdot\,) = \phi_\nu(\,\cdot\,; \widetilde{\phi}_1,\vec m)$ \Comment{$\phi_2$ is given by the single layer potential \eqref{eq:single_layer_pot}}
\phase{Automatic differentiation:}
\State $\widetilde{\vec h}_d = -\nabla (\widetilde{\phi}_1 + \widetilde{\phi}_2)$ \Comment{compute stray field with forward mode AD}
\end{algorithmic}
\end{algorithm}
To compute the magnetic induction, one would use \eqref{eq:A1-poisson} and \eqref{eq:A2-single_layer_pot} instead. If one uses a more complicated network structure than an ELM, the computation of $\phi_1$ needs to be replaced with the respective minimization problem. Alternatively to solving the single layer potential, one could compute the Dirichlet boundary data to train a PINN or an ELM to solve the respective Laplace problem \eqref{eq:phi2-interface}.

Section \ref{sec:pinns} explains in detail how to solve a problem like \eqref{eq:phi1-poisson} with PINNs. Especially the hard constraint formulation $\phi_1(\vec x) = \ell(\vec x) \hat\phi_1(\vec x)$ (see section \ref{sec:hard_constraints}) is of interest. Here, $\ell$ is an approximate distance function (ADF), normalized to first order, i.e., $-\left.\nabla\ell\right|_{\partial\Omega}=\left.\vec n\right|_{\partial\Omega}$, and $\hat\phi_1$ is a neural network model. With this ansatz, equation \eqref{eq:single_layer_pot} simplifies to
\begin{equation}\label{eq:single_layer_pot_improved}
    \phi_2(\vec x) = \frac{1}{4\pi}\int_{\partial\Omega} \frac{(\vec m\cdot \vec n + \hat\phi_1)(\vec y)}{\|\vec x - \vec y \|}\, \mathrm{d}s(\vec y).
\end{equation}
In this work, this integral is solved numerically. Numerical integration can either be performed with a Monte Carlo approach, as in \cite{schaffer2023physics}, or with integration over surface elements forming a partition \cite{schaffer2024constraint}. The latter method has the advantage of being more cost-efficient and accurate. Also, it allows for a pre-computation step, which significantly reduces computational work during the actual computation. 

\subsection{Evaluation of the energy bounds on the finite domain}\label{sec:brown-revisited}
For our numerical experiments, we are interested in the evaluation of the lower energy bound \eqref{eq:phi-energy-func} and the upper energy bound \eqref{eq:A-energy-func}.
In principle, a truncation approach could be used to evaluate the respective whole space integrals, but this still requires a very big domain outside the magnet and further introduces a truncation error. In the previous section, we used the splitting ansatz of Garcia-Cervera and Roma to derive a solution to the whole space transmission problems \eqref{eq:stray-field} and \eqref{eq:b-field}. This splitting ansatz can also be used for the computation of the energy bounds without considering the external domain $\overline\Omega^c$.

Two important properties can be used to derive a version of the lower bound \eqref{eq:bounds} on the finite domain:
\begin{lem}[\cite{exl2018magnetostatic}]\label{lem:props}
Let $\phi_1\in H^1_0(\Omega)$ and $\phi_2 \in \left(H^1(\Omega)\times H^1_{loc}(\overline\Omega^c)\right)$ solving the Laplace equation. Then we have\\[0.1cm]
(i) The gradients $\nabla \phi_1$ and $\nabla \phi_2$ are $L^2$-orthogonal, that is,
\begin{align}
\int_{\Omega} \nabla \phi_1 \cdot \nabla \phi_2 \, \mathrm{d}\vec x =  0.
\end{align}
(ii) Let $ \phi_2 $ be as in \eqref{eq:single_layer_pot}, then the $L^2$-norm of $\nabla \phi_2$ can be expressed as a double surface integral
\begin{align}\label{eqn:inner_g}
 \int_{\mathbb{R}^3}  \|\nabla \phi_2\|^2\,\mathrm{d}\vec x = \int_{\partial\Omega} (\vec m\cdot \vec n - D_{\vec n}\phi_1) \phi_2\, \mathrm{d}s(\vec x).   
\end{align}
\hfill $\square$
\end{lem}

Using the splitting of \textit{Garcia-Cervera and Roma} $\phi = \phi_1 + \phi_2$ with $\phi_1\in H^1_0(\Omega)$ and $\phi_2 = \phi_\nu(\vec x;\,\phi_1,\vec m)$ being the single layer potential \eqref{eq:single_layer_pot}, as well as property $(i)$ and $(ii)$ from Lemma~\ref{lem:props} and applying it to the energy functional \eqref{eq:phi-energy-func} yields
\begin{equation}\label{eq:lowerbound-finite}
   e_\phi(\phi_1;\,\vec m) = -\int_{\Omega} \|\nabla \phi_1\|^2 \, \mathrm{d}\vec x + 2 \int_{\Omega} \vec m \cdot(\nabla \phi_1 + \nabla \phi_2) \, \mathrm{d}\vec x - \int_{\partial\Omega} (\vec m\cdot \vec n - D_{\vec n}\phi_1) \cdot \phi_2\, \mathrm{d}s(\vec x).
\end{equation}
Note that $\phi_1$ does not satisfy the Dirichlet Poisson problem \eqref{eq:phi1-poisson} here, but is a free function.
The surface integral can be transformed into a volume integral using the divergence theorem, together with property $(i)$,
\begin{align}\label{eq:phi1-double-surface-int}
    \int_{\partial\Omega} (\vec m\cdot \vec n - D_{\vec n}\phi_1) \cdot \phi_2\, \mathrm{d}s(\vec x) = 
    \int_{\Omega} (\nabla\cdot\vec m) \phi_2 + \vec m\cdot \nabla \phi_2 - (\Delta \phi_1)\phi_2 \, \mathrm{d}\vec x
\end{align}

Lemma~\ref{lem:props} can be restated for the vector potential, consequently, the upper bound \eqref{eq:A-energy-func}, with $\vec A_2 = \vec A_\nu(\vec x;\,\vec A_1,\vec m)$ as in \eqref{eq:A2-single_layer_pot}, can be written as
\begin{equation}\label{eq:upperbound-finite}
\begin{aligned}
   e_{\vec A}(\vec A_1;\vec m) \,& = \int_{\Omega} \|\boldsymbol{m}\|^2\,\mathrm{d}\vec x + \int_{\Omega} |\nabla \vec A_1|^2\,\mathrm{d}\vec x - 2 \int_{\Omega} \vec m\cdot \big(\nabla \times (\vec A_1+\vec A_2)\big) \, \mathrm{d}\vec x \\
    &+ \int_{\partial\Omega} (\vec m\times \vec n - D_{\vec n}\vec A_1) \cdot \vec A_2\, \mathrm{d}s(\vec x).
\end{aligned}
\end{equation}
Again, it is possible to write the surface integral as volume integral like
\begin{equation}\label{eq:A1-double-surface-int}
    \int_{\partial\Omega} (\vec m\times \vec n - D_{\vec n}\vec A_1) \cdot \vec A_2\, \mathrm{d}s(\vec x) = 
    -\int_{\Omega} (\nabla\times\vec m)\cdot \vec A_2 - \vec m\cdot \nabla \times\vec A_2 + (\Delta \vec A_1)\cdot\vec A_2 \, \mathrm{d}\vec x.
\end{equation}
Especially for numerical integration of the single layer potential it is useful not to integrate over the surface, but the volume instead since singularities are easier to control. 

Further, for the unique solution of \eqref{eq:phi1-poisson}, $\phi_d = \phi_1 + \phi_2$ defines the unique solution of \eqref{eq:stray-field}. The field $-\nabla (\phi_1 + \phi_2)$ defines the stray field $\vec h_d$, where the energy $e_d(\vec m) = - \int_{\Omega} \vec m \cdot \vec h_d \,\mathrm{d}\vec x$ can also be expressed as \cite{exl2018magnetostatic} 
\begin{align}\label{eqn:energy_new}
    e_d(\vec{m}) = \int_{\Omega} \|\nabla\phi_1\|^2\,\mathrm{d}\vec x + \int_{\partial\Omega} (\vec m\cdot \vec n - D_{\vec n}\phi_1) \cdot \phi_2\, \mathrm{d}s(\vec y).
\end{align}

\section{Micromagnetic Energy Minimization}\label{ch:energy-minimization}

Micromagnetic energy minimization aims to minimize the Gibbs free energy \cite{exl2020micromagnetism} of the magnet $\Omega \subset \mathbb R^3$ which is given in reduced units as

\begin{equation}\label{eq:totenergy}
\begin{aligned}
    e = &\frac{1}{2}e_d + \underbrace{\int_\Omega - \vec m \cdot \vec h_{ext}\mathrm{d}\vec x}_{e_{zee}} + \underbrace{\int_\Omega Q (1 - (\vec m \cdot \vec a)^2)\mathrm{d}\vec x}_{e_a} + \underbrace{\int_\Omega \widetilde{A}_{ex}\|\nabla\vec m\|^2_F\mathrm{d}\vec x}_{e_{ex}},
\end{aligned}
\end{equation}
with the four fundamental energy terms: magnetostatic energy $e_d$ as in \eqref{eq:demag-energies}, Zeeman energy $e_{zee}$, an\-iso\-tropy energy $e_a$ and exchange energy $e_{ex}$. The reduced energy has units $[e]=\unit{m^3}$. The energy functional in reduced units is derived by dividing the Gibbs free energy with the energy density $K_m:=\mu_0M_s^2$ ($[\mu_0M_s^2]=\unit{J / m ^ 3}$). The quantity $\mu_0=4\pi\qty{e-7}{Tm/A}$ is the vacuum permeability and $M_s$ (in units of $\unit{A/m}$) is the saturation magnetization. Therefore, the reduced uniaxial anisotropy constant is denoted with $Q:=K_u / K_m$ with $K_u$ being the uniaxial anisotropy constant and $\vec a$ is the easy axis of the magnet. Further, $\widetilde{A}_{ex} := A_{ex} / K_m$ is the reduced exchange stiffness constant. One can for instance apply \textit{LaBonte's method} to minimize the energy and perform a curvilinear steepest descent optimization with the projected gradient of the energy to find a local minimum \cite{exl2014labonte}, or variants of nonlinear conjugate gradient methods \cite{fischbacher2017nonlinear,exl2019preconditioned}. For a nonlinear PINN ansatz, optimization is more intricate.

\subsection{Optimization of the Gibbs free energy functional}
Let the magnetization be parameterized by the model $\widetilde{\vec m}(\;\cdot\;;\vec\theta)$, it is possible to use AD to compute the gradient $\nabla_{\vec\theta}e$ of the energy functional with respect to the model parameters $\vec\theta$ and use (stochastic) gradient descent to update the model parameters. Also higher order methods can be applied. However, two major problems arise. The first problem is the satisfaction of the unit norm constraint, $\|\vec m\| = 1$, which will be explained later on. The second one arises from the non-local interaction due to the stray field. 

When approximating the integral in \eqref{eq:totenergy}, the integrand needs to be computed pointwise. The stray field is a function of the magnetization, $\vec h_d = \vec h_d(\;\cdot\;;\widetilde{\vec m}, \vec \theta)$, thus, computing the self energy for instance with
\begin{equation}
    e_d(\vec \theta) = -\int_\Omega \widetilde{\vec m}(\vec x;\vec\theta) \cdot \vec h_d(\vec x;\widetilde{\vec m}, \vec \theta) \mathrm{d} \vec x
\end{equation}
requires AD through the stray field solver. This can be done with ease with modern AD libraries but comes with a big computational burden. Using the splitting ansatz of Garcia-Cervera and Roma in Section~\ref{sec:garcia-cervera-roma}, $\vec h_d = -\nabla(\phi_1 + \phi_2)$, and for instance a PINN to compute the approximate solution to $\widetilde{\phi}_1$ and the single layer potential \eqref{eq:single_layer_pot} $\widetilde{\phi}_2 =\phi_\nu(\;\cdot\;;\widetilde{\phi}_1,\widetilde{\vec m})$ makes AD very expensive and prohibitive. If $\widetilde{\phi}_1$ is computed via an optimization process, it would require implicit AD \cite{blondel2022efficient} of the solver. Further, AD of the single layer potential requires AD of the integral operator, which is expensive for large domains or grain structure models. To our luck, it is not necessary to compute the gradient of the stray field with respect to the model parameters $\vec\theta$, since the stray field is a linear function of the magnetization, $\vec h_d(\,\cdot\,;\widetilde{\vec m},\vec\theta) = \vec D(\,\cdot\,)\widetilde{\vec m}(\,\cdot\,;\vec\theta)$, with $\vec D$ being the constant demagnetization tensor which is only dependent on the geometry. Hence, the partial derivative of $e_d$ with respect to some parameter of the model $\vec\theta_i$ is given by
\begin{equation}\label{eq:stray_field_grad}
    \frac{\partial e_d}{\partial \vec\theta_i} 
    = -\frac{\partial}{\partial \vec\theta_i} \int_\Omega \widetilde{\vec m} \cdot \vec D \widetilde{\vec m}\, \mathrm{d}\vec x 
    = -2\,\int_\Omega \vec h_d \frac{\partial \widetilde{\vec m}}{\partial \vec\theta_i} \, \mathrm{d}\vec x.
\end{equation}

Using a first order optimization method such as Stochastic Gradient Descent (SGD), only the gradient is required to update the model parameters. Therefore, one can compute the stray field pointwise, before evaluating the gradient $\nabla_{\vec \theta} (e_d + e_{zee} + e_a + e_{ex})$. Note that the factor $\frac{1}{2}$ is missing before the self energy term. 
This approach results in Algorithm~\ref{alg:energy-minimization1}, where
\begin{equation}\label{eq:integrand-energy-min}
    q(\vec x, \vec h; \vec\theta) = -\widetilde{\vec m}(\vec x;\vec\theta) \cdot \vec h - \widetilde{\vec m}(\vec x;\vec\theta) \cdot \vec h_{ext}(\vec x) 
    + Q(1 - (\widetilde{\vec m}(\vec x;\vec\theta) \cdot \vec a)^2) + \widetilde{A}_{ex}\|\nabla \widetilde{\vec m}(\vec x;\vec\theta)\|_F^2
\end{equation}
is the corresponding integrand of $e_d + e_{zee} + e_a + e_{ex}$, and $\vec h\in \mathbb{R}^3$ is the pointwise stray field. All other terms of \eqref{eq:totenergy} can be computed pointwise using automatic differentiation. All this also holds for the vector potential and the magnetic flux density $\vec b_d$. The algorithm can be formulated analogous.

\begin{algorithm}
\caption{Energy minimization with first order methods}
\label{alg:energy-minimization1}
\begin{algorithmic}
\Require Iteration limit $N_{\vec m}$
\For{$1,\dots,N_{\vec m}$}
\State 1. Draw batch $\vec X\in\Omega$ from the computational domain
\State 2. Compute $\vec h_{d,i} = \vec h_d(\vec x_i)$ for all $\vec x_i \in \vec X$. \Comment{e.g. with Algorithm~\ref{alg:elm-stray-field}}
\State 3. Compute updates using AD
\begin{equation*}
\vec g = \nabla_{\vec \theta} \left(\frac{1}{|\vec X|}\sum_{\vec x_i \in \vec X} q(\vec x_i; \vec h_{d,i}, \vec \theta)\right)
\end{equation*}
\State 4. Update the parameters $\vec \theta$ with $\vec g$ using some first order optimizer
\EndFor
\end{algorithmic}
\end{algorithm}
This algorithm also perfectly fits into a conditional framework, where the magnetization model includes additional conditional parameters. They can be incorporated into the model by appending them to the input vector of the neural network \cite{kovacs2022conditional}. In that case, one can draw a full batch or minibatch of the domain $\Omega$ as well as a minibatch of the conditional domain and compute the stray field for each conditional sample. The ADF which is used for computation of the scalar potential can also be parameterized, accounting for different geometries. However, one should note that the sample space grows exponentially with the dimension and therefore, the computational load of evaluating \eqref{eq:totenergy} over all conditional parameters also grows exponentially.
Note that the computation of the magnetostatic fields can also be done with any other method. For instance, \textit{MagTense} \cite{bjork2021magtense} is one such framework.

The remaining problem of Algorithm~\ref{alg:energy-minimization1} is that it does not deal with the satisfaction of the unit norm constraint. In the following, two different approaches are introduced. The first one uses a penalty framework and therefore changes the algorithmic framework, while the second one imposes the constraints on the model level and can apply Algorithm~\ref{alg:energy-minimization1} as given.

\subsection{Penalty framework}
A penalty framework can be used to account for the unit norm constraint \cite{schaffer2023physics}, i.e.,
\begin{equation}\label{eq:totenergy-penalty}
\begin{aligned}
    e_\alpha(\vec\theta) = e(\vec\theta) + \alpha \int_\theta(\|\vec m(\vec x;\,\vec\theta)\| - 1)^2\,\mathrm{d}\vec x \rightarrow \mathrm{min}_{{\vec\theta}}
\end{aligned}
\end{equation}
with the penalty parameter $\alpha$. Starting with a low value for the penalty parameter, after minimization of the objective, it is increased for each consecutive optimization. The constrained is fulfilled if $\alpha\rightarrow \infty$. However, this caused the problem to be more ill-conditioned \cite{nocedal1999numerical}. Therefore, one needs to stop the algorithm if the penalty parameter becomes prohibitively large. Algorithm~\ref{alg:penalty-framework} gives a pseudocode for such a framework. It builds on Algorithm~\ref{alg:energy-minimization1} and adds an outer loop to control the size of the penalty parameter.
\begin{algorithm}
\caption{Penalty framework for Gibbs free energy minimization}
\label{alg:penalty-framework}
\begin{algorithmic}
\Require Iteration limits $N_{\alpha}$, $N_{\vec m}$; initial penalty $\alpha_0$; increase factor $\zeta > 1$
\State $\alpha \gets \alpha_0$
\For{$1,\dots,N_{\alpha}$}
\For{$1,\dots,N_{\vec m}$}
\State 1. Draw batch $\vec X\in\Omega$ from the computational domain
\State 2. Compute $\vec h_{d,i} = \vec h_d(\vec x_i)$ for all $\vec x_i \in \vec X$. \Comment{e.g. with Algorithm~\ref{alg:elm-stray-field}}
\State 3. Compute updates of penalized objective using AD
\begin{equation*}
\qquad\vec g = \nabla_{\vec \theta} \left(\frac{1}{|\vec X|}\sum_{\vec x_i \in \vec X} q(\vec x_i; \vec h_{d,i}, \vec \theta) + \alpha (\|\widetilde{\vec m}(\vec x_i;\vec\theta)\| - 1)^2\right) 
\end{equation*}
\State 4. Update the parameters $\vec \theta$ with $\vec g$ using some first order optimizer.
\EndFor
\State $\alpha \gets \zeta\alpha$ \Comment{increase penalty parameter}
\EndFor
\end{algorithmic}
\end{algorithm}

\subsection{Cayley transform}\label{sec:cayley}
Another possibility to account for the unit norm constraint is the utilization of the Cayley-transform \cite{krishnaprasad2001cayley}. This assures that the model output is given on the Lie algebra of the $SO(3)$ rotation group, and this rotation can be applied to some initial magnetization $\vec m_0$ \cite{schaffer2024constraint}. Specifically, we consider the isomorphism between $\mathbb R^3$ and the Lie algebra $so(3)$ of $3\times 3$ skew symmetric matrices
\begin{equation}
    \vec p=\begin{pmatrix}
        p_1\\p_2\\p_3
    \end{pmatrix} \mapsto
    \hat{\vec p} = \begin{pmatrix}
        0 & -p_3 & p_2\\
        p_3 & 0 & -p_1\\
        -p_2 & p_1 & 0
    \end{pmatrix},
\end{equation}
and use the Cayley transform
\begin{equation}
    \mathrm{cay}(\vec p) = (I - \hat{\vec p})^{-1}(I + \hat{\vec p})
\end{equation}
to describe the rotation of the initial magnetization. For some input $\vec x\in\Omega$, 
a neural network $\mathcal{N}:\Omega \times \mathbb R^M \rightarrow \mathbb R^3,(\vec x, \vec \theta) \mapsto \mathcal{N}(\vec x; \vec \theta)$ with parameters $\vec\theta \in \mathbb R^M$ and some initial magnetization $\vec m_0$, the hard constraint model takes the form
\begin{equation}\label{eq:mag_model}
    \widetilde{\vec m}(\vec x, \vec \theta) = \mathrm{cay}\left(\mathcal N(\vec x;\vec\theta)\right)\vec m_0(\vec x).
\end{equation}
This architecture allows energy minimization of \eqref{eq:totenergy} using Algorithm~\ref{alg:energy-minimization1} without the need of any framework for constrained optimization.

\subsection{Optimization of the energy bounds on the finite domain}\label{sec:opt_bounds}
Up to this point, only first order optimization method were considered. If using second order optimizers, not only the gradient but also the objective value is important, and they need to be consistent. This can be problematic for implementation of second order methods, since one need to consider the objective $e = \frac{1}{2}e_d + e_{zee} + e_a + e_{ex}$ but with gradients given by $\nabla_{\vec \theta} (e_d + e_{zee} + e_a + e_{ex})$, assuming the stray field is independent of the magnetization and the gradient of the self energy is given by \eqref{eq:stray_field_grad}. In a first version, it was thought that the objective $e_d + e_{zee} + e_a + e_{ex}$ can replace $e$ and AD of the stray field can simply be prevented on a code level with the \texttt{JAX} directive \texttt{stop\_gradient}. However, this does not work for second order methods, since the gradient corresponds to $e$ but the value does not. This issue can probably easily be resolved on a code level, but the study on energy bounds opened a new viewpoint. It turns out, that the minimization of $\bar e = e_d + e_{zee} + e_a + e_{ex}$, without the factor $\frac{1}{2}$ for the self energy, corresponds to minimization of the lower bound $e_{\phi}$ in \eqref{eq:phi-energy-func} with respect to the magnetization. In essence, this means that one can consider the value of $\bar e$ instead of $e$. Therefore, in this section, the joint optimization of $\vec m$ and the energy bounds is explained and results in a sophisticated alternating minimization algorithm.\newline

Equation \eqref{eq:lowerbound-finite} and \eqref{eq:upperbound-finite} give a computable formulation of the energy bounds on the finite domain and could be tackled, for instance, with Monte Carlo integration. However, we were not able to use those bounds directly for energy minimization in a physics-informed setting. It would require AD over the single layer potential, and minimization of the energy functional with PINNs does not seem to be numerically sound.

When training a physics-informed model like a PINN, in essence, we are only interested in the updates of the model parameters. 
Considering again the lower bound \eqref{eq:phi-energy-func} of the self energy
\begin{align*}
   e_\phi(\phi; \,\vec m) = -\int_{\mathbb{R}^3} \|\nabla \phi\|^2 \, \mathrm{d}\vec x + 2\int_{\Omega} \vec m \cdot \nabla \phi\, \mathrm{d}\vec x.
\end{align*}
The goal is the maximization of this functional with respect to the scalar potential $\phi$, but minimize it with respect to the magnetization $\vec m$. Hence, this leads to a min-max problem.
Computing the variational derivative with a variation $\delta \phi$ with $[\delta \phi]=0$ across the boundary $\partial\Omega$ yields
\begin{equation}
\begin{aligned}
    \mathrm{d}_{\phi} e_\phi(\phi; \,\vec m) = 2\Bigl( &\int_{\mathbb{R}^3} (\Delta\phi)\delta\phi \, \mathrm{d}\vec x - \int_{\Omega} (\nabla\cdot\vec m)\delta\phi\, \mathrm{d}\vec x \\ 
    &+ \int_{\partial\Omega} ([D_n\phi] + \vec m \cdot \vec n)\delta\phi\, \mathrm{d}s(\vec x)\Bigl).
\end{aligned}   
\end{equation}
The Fréchet derivatives, and hence the local updates, are given by
\begin{align}
    \frac{\partial}{\partial \phi} e_\phi(\phi;\vec m) &\, = 2\, (\Delta \phi - \nabla\cdot\vec m) &\text{in} \,\, \Omega,\\ 
    \frac{\partial}{\partial \phi} e_\phi(\phi;\vec m) &\, = 2\, \Delta \phi &\text{in} \,\, \overline\Omega^c,\\
    \frac{\partial}{\partial \phi} e_\phi(\phi;\vec m) &\, = 2\, ([D_n \phi] + \vec m \cdot \vec n) &\text{on} \,\, \partial\Omega,
\end{align}
and variation of $\vec m$ yields
\begin{align}\label{eqn:gradm}
    \frac{\partial}{\partial \vec m} e_\phi(\phi;\boldsymbol{m}) &\, = 2\,( \nabla\phi) &\text{in} \,\, \Omega.
\end{align}
Decomposing $\phi$ into $\phi = \phi_1 + \phi_2$  with $\phi_1$ zero at the boundary and extended with zero in $\overline\Omega^c$ and $\phi_2$ satisfying Laplace equation in $\Omega \cup \overline\Omega^c$, continuous across $\partial\Omega$ and fulfilling the jump in the normal derivative as $[D_n \phi_2] = D_n \phi_0 - \vec m\cdot\vec n$, we arrive at the following gradients
\begin{align}
    \frac{\partial}{\partial \phi} e_\phi(\phi;\vec m) &\, = 2\, (\Delta \phi_1 - \nabla\cdot\vec m)  &\text{in} \,\, \Omega,\\ 
    \frac{\partial}{\partial \phi} e_\phi(\phi;\vec m) &\, = 0  &\text{in} \,\, \overline\Omega^c,\\
    \frac{\partial}{\partial \phi} e_\phi(\phi;\vec m) &\, = 0  &\text{on} \,\, \partial\Omega.
\end{align}
Therefore, only $\phi_1$ but not $\phi_2$ needs to be updated to maximize the energy functional with respect to $\phi_1$.
Let $\phi_\nu(\,\cdot\,;\,\phi_1,\vec m)$ be the single layer potential \eqref{eq:single_layer_pot}. If $\phi_1$ is parameterized by a hard constraint PINN $\widetilde\phi_1(\,\cdot\,;\vec\omega)$ (see section \ref{sec:hard_constraints}) with model parameters $\vec\omega$ and $\vec m$ is parameterized by the model $\widetilde{\vec m}(\,\cdot\,;\,\vec\theta)$ with parameters $\vec\theta$, we can consider the following objectives
\begin{align}
    \mathcal{L}_{\phi_1}(\vec\omega, \vec\theta) &= \int_\Omega |\Delta\widetilde\phi_1(\vec x;\,\vec\omega) - \nabla\cdot\widetilde{\vec m}(\vec x;\,\vec\theta)|^2 \mathrm{d}\vec x \label{eq:phi1-objective}\\
    e_\phi(\vec\omega, \vec\theta, \bar{\vec\theta}) &= -\int_{\mathbb{R}^3} \|\nabla \widetilde{\phi}(\vec x;\,\vec\omega,\bar{\vec\theta})\|^2 \, \mathrm{d}\vec x + 2\,\int_{\Omega} \nabla \widetilde{\phi}(\vec x;\,\vec\omega,\bar{\vec\theta})\cdot \widetilde{\vec m}(\vec x;\,\vec\theta)\,\mathrm{d}\vec x \label{eq:m-objective},
\end{align}
with $\widetilde{\phi}(\vec x;\,\vec\omega,\bar{\vec\theta}) = \widetilde\phi_1(\vec x;\,\vec\omega) + \phi_\nu(\vec x;\widetilde\phi_1(\,\cdot\,;\vec\omega), \widetilde{\vec m}(\,\cdot\,;\,\bar{\vec\theta}))$. 
The update for $\phi_1$ is replaced by the update resulting from the strong residual of \eqref{eq:phi1-poisson} and $\bar{\vec\theta}$ is independent of $\vec\theta$.
The derivatives with respect to some model parameters $\vec\omega_i$ and $\vec\theta_j$ are given by
\begin{align}
    \frac{\partial}{\partial \vec\omega_i}\mathcal{L}_{\phi_1}(\vec\omega, \vec\theta) &= 2\,\int_\Omega \left(\Delta\widetilde\phi_1(\vec x;\,\vec\omega) - \nabla\cdot\widetilde{\vec m}(\vec x;\,\vec\theta)\right)\frac{\partial\Delta\widetilde\phi_1(\vec x;\,\vec\omega)}{\partial \vec\omega_i} \mathrm{d}\vec x \\
    \frac{\partial}{\partial \vec\theta_j}e_\phi(\vec\omega, \vec\theta, \bar{\vec\theta}) &= 2\,\int_{\Omega} \nabla\widetilde{\phi}(\vec x;\,\vec\omega,\bar{\vec\theta})\cdot \frac{\partial \widetilde{\vec m}(\vec x;\,\vec\theta)}{\partial \vec\theta_j}\,\mathrm{d}\vec x.\label{eq:e_phi_grad}
\end{align}
Both \eqref{eq:phi1-objective} and \eqref{eq:m-objective} need to be minimized. 
Exactly the same can be done for the vector potential $\vec A = \vec A_1 + \vec A_2$ with parametrization $\widetilde{\vec A_1}(\,\cdot\,;\vec\omega)$ with the objectives
\begin{align}
    \mathcal{L}_{\vec A_1}(\vec\omega, \vec\theta) &= \int_\Omega \|\Delta\widetilde{\vec A}_1(\vec x;\,\vec\omega) + \nabla\times\widetilde{\vec m}(\vec x;\,\vec\theta)\|^2 \mathrm{d}\vec x \label{eq:A1-objective}\\
\begin{split}
    e_{\vec A}(\vec\omega, \vec\theta, \bar{\vec\theta}) &= \int_{\Omega} \|\widetilde{\vec m}(\vec x;\,\vec\theta)\|^2\,\mathrm{d}\vec x + \int_{\mathbb{R}^3} \|\nabla \widetilde{\vec A}(\vec x;\,\vec\omega,\bar{\vec\theta})\|^2 \, dx \\
    &- 2\,\int_{\Omega} \left(\nabla\times\widetilde{\vec A}(\vec x;\,\vec\omega,\bar{\vec\theta})\right) \cdot \widetilde{\vec m}(\vec x;\,\vec\theta)\,\mathrm{d}\vec x \label{eq:m-A1-objective},    
\end{split}
\end{align}
with $\widetilde{\vec A}(\vec x;\,\vec\omega,\bar{\vec\theta}) = \widetilde{\vec A}_1(\vec x;\,\vec\omega) + \vec A_\nu\left(\vec x;\widetilde{\vec A}_1(\,\cdot\,;\vec\omega), \widetilde{\vec m}(\,\cdot\,;\,\bar{\vec\theta})\right)$
and $\vec A_\nu$ as in \eqref{eq:A2-single_layer_pot}. 
Again, only the last term is relevant for the optimization of $\widetilde{\vec m}$.

Comparing \eqref{eq:e_phi_grad} with \eqref{eq:stray_field_grad} and setting $h_d = -\nabla\widetilde{\phi}$, we see that the gradients match. 
The magnetostatic energy $e_d$ in \eqref{eq:totenergy} can be replaced by \eqref{eq:m-objective} or \eqref{eq:m-A1-objective} which yields the objectives
\begin{align}
    \hat e(\vec\omega, \vec\theta, \bar{\vec\theta}) &= \frac{1}{2}e_{\phi}(\vec\omega, \vec\theta, \bar{\vec\theta}) + e_{zee}(\vec\theta) + e_a(\vec\theta) + e_{ex}(\vec\theta)\label{eq:totenergy_lower} \\
    \check e(\vec\omega, \vec\theta, \bar{\vec\theta}) &= \frac{1}{2}e_{\vec A}(\vec\omega, \vec\theta, \bar{\vec\theta}) + e_{zee}(\vec\theta) + e_a(\vec\theta) + e_{ex}(\vec\theta) \label{eq:totenergy_upper}
\end{align}

Those formulations still include a whole space integral. The evaluation on the finite domain is explained in Section~\ref{sec:brown-revisited}. However, since $e_{\phi}$ and $e_{\vec A}$ are minimized with respect to $\vec\theta$ but not $\bar{\vec\theta}$, the whole space integral can be treated as constant and be removed from the objective.
Algorithm~\ref{alg:energy-minimization2} shows the minimization of the energy functional using an arbitrary first or second order optimizer, assuming that the unit norm constraint is satisfied via hard constraint formulation. This algorithm is a generalization of Algorithm~\ref{alg:energy-minimization1}, where the stray field is not computed explicitly.

\begin{algorithm}
\caption{Energy minimization with lower energy bound for the self-energy}
\label{alg:energy-minimization2}
\begin{algorithmic}
\Require $\hat e$, $\vec\omega$, $\vec\theta$, tolerance $\epsilon_{\vec m}$ 
\Ensure $\|\nabla_{\vec\theta}\hat e(\vec\omega, \vec\theta,\bar{\vec\theta})\| < \epsilon_{\vec m}$
\State $\bar{\vec\theta} \gets \vec\theta$
\Loop
\State 1. Find approximate solution $\widetilde{\phi}_1$ with parameters $\vec \omega$ \Comment{ELM or PINN solution}
\State 2. Perform minimization step with objective $\hat e(\vec \omega, \vec\theta, \bar{\vec\theta})$ with respect to $\vec\theta$
\State $\bar{\vec\theta} \gets \vec\theta$
\EndLoop
\end{algorithmic}
\end{algorithm}

When using a hard constraint PINN ansatz for $\widetilde{\phi}_1$ and $\widetilde{\vec m}$, an alternating optimization scheme can be used for energy minimization. First, the solution to $\widetilde{\phi}_1$ is replaced with the respective PINN minimization problem, and second, we observe that more than one steps can be performed for each iteration. This leads to Algorithm~\ref{alg:phi1-ed-min}.

\begin{algorithm}
\caption{Alternating scheme for energy minimization}
\label{alg:phi1-ed-min}
\begin{algorithmic}
\Require $\hat e$, $\mathcal{L}_{\phi_1}$, $\vec\omega$, $\vec\theta$, tolerances $\epsilon_{\vec m}$ and $\epsilon_{\phi_1}$, iteration limits $N_{\vec m}$ and $N_{\phi_1}$
\Ensure $\|\nabla_{\vec\omega}\mathcal{L}_{\phi_1}(\vec\omega, \vec\theta)\| < \epsilon_{\phi_1}$ and $\|\nabla_{\vec\theta}\hat e(\vec\omega, \vec\theta,\bar{\vec\theta})\| < \epsilon_{\vec m}$
\State $\bar{\vec \theta} \gets \vec \theta$
\Loop
    \For{$1,\dots,N_{\phi_1}$}
        \State $\vec\omega \gets$ Perform minimization step for $\mathcal{L}_{\phi_1}(\vec\omega, \vec\theta)$ w.r.t $\vec\omega$
        \If{$\|\nabla_{\vec\omega}\mathcal{L}_{\phi_1}(\vec\omega, \vec\theta)\| < \epsilon_{\phi_1}$}
            \Break
        \EndIf
    \EndFor
    \For{$1,\dots,N_{\vec m}$}
        \State $\vec\theta \gets$ Perform minimization step for $\hat e(\vec\omega, \vec\theta,\bar{\vec\theta})$ w.r.t $\vec\theta$
        \If{$\|\nabla_{\vec\theta}\hat e(\vec\omega, \vec\theta, \bar{\vec\theta})\| < \epsilon_{\vec m}$}
            \Break
        \EndIf
    \EndFor
    \State $\bar{\vec \theta} \gets \vec \theta$
\EndLoop
\end{algorithmic}
\end{algorithm}

All the introduced algorithms can equally be applied with the vector potential and the upper bound instead. Minimization of the vector potential can be performed with the same algorithm by just replacing $\mathcal{L}_{\phi_1}$ with $\mathcal{L}_{\vec A_1}$ and by using $\check e$. If $\phi_1$ or $\vec A_1$ is modeled with a hard constrained ELM ansatz, with a precomputed solution operator, the inner loop to update $\vec\omega$ can be replaced with the solution of the ELM. Further, if the unit norm constraint is modelled with a penalty framework, one needs to add the respective penalty term and gradually increase the penalty parameter during optimization.

The proposed energy minimization method can be applied for the computation of a hysteresis curve. The hysteresis depends on the history of previous magnetization states as a function of the external field. 
Incorporating the external field as conditional parameter can therefore be difficult and might lead to an optimization problem which is not well-posed. Moreover, a simple continuous neural network model cannot approximate discontinuities well and the energy functional needs to be properly minimized for each external field. Therefore, a better choice of optimization might be the application of a full batch second order optimizer such as the \textit{Trust Region} method and increase/decrease the external field step-by-step.

To minimize grain structure models, one could use a separate small scale PINN to model the magnetization for each grain. Further, and ELM or PINN for each grain would be required to model scalar or vector potential. The training for the scalar or vector potential model can be performed in parallel on many computational units. The same holds true for the magnetization. Only computation of the single layer potential \eqref{eq:single_layer_pot} needs synchronization. Therefore, better evaluation strategies for the single layer potential are of particular importance.

\section{Results}
This section includes results of scalar and vector potential for simple domains with fixed magnetization, and also computes the energy bounds \eqref{eq:phi-energy-func} and \eqref{eq:A-energy-func}, with the respective formulas, \eqref{eq:lowerbound-finite} and \eqref{eq:upperbound-finite}, introduced in Section~\ref{sec:brown-revisited}. Algorithm~\ref{alg:elm-stray-field} is used for the computation of the scalar potential and the respective stray field. An analogous version of the algorithm is used for the vector potential. Further, in Section~\ref{sec:flower-state}, $\phi_1$ and $\vec A_1$ are also modelled via PINNs to have a comparison to the ELM solution. Later on, we use Algorithm~\ref{alg:energy-minimization2} and \ref{alg:phi1-ed-min} for full 3d Gibbs free energy minimization to compute the demagnetization curve of a spherical and cubic particle. In \cite{schaffer2024constraint} and \cite{schaffer2023physics} Algorithm~\ref{alg:energy-minimization1} and Algorithm~\ref{alg:penalty-framework} were applied to compute the solution to the NIST $\mu$MAG Standard Problem \#3 \cite{mumag3}.

\subsection{Outward magnetized sphere}
As a first test for the computation of the micromagnetic self energy and Algorithm~\ref{alg:elm-stray-field}, we considered a spherical magnetic particle with radius $R=1$ and outward facing magnetization $\vec m(\vec x) = \vec x / \|\vec x\|$ centered at the origin. The exact solution to the scalar potential is given by
\begin{equation}
\begin{aligned}
\phi_d = \bigg\{
\begin{array}{ll}
\|\vec x\| - 1 & \text{in}\; \Omega\\ 
0 & \text{in}\; \overline\Omega^c.\\ 
\end{array}
\end{aligned}
\end{equation}
For this example, the scalar potential is zero at the boundary, hence $\phi_d = \bar{\phi}_1$ and $\bar{\phi}_2=0$. The energy is therefore $e_d=\frac{4}{3}\pi$. However, $\phi_2$ was also included in the numerical test since errors in $\phi_1$ will propagate. The solution of $\phi_1$ was computed with a hard constraint ELM with the ADF $\ell(\vec x) = (R^2 - \|\vec x\| ^ 2) / 2R)$, i.e.,
\begin{equation}\label{eq:hard-constraint-elm-sphere}
    \widetilde{\phi}_1(\vec x;\,\vec\beta) = \ell(\vec x) \hat{\phi}_1(\vec x;\,\vec\beta),
\end{equation}
where $\hat{\phi}_1$ is an ELM model. Therefore, the full model for the scalar potential is given by 
\begin{equation}
    \phi_{ELM}(\vec x;\,\vec \beta) = \widetilde{\phi}_1(\vec x;\,\vec\beta) + \phi_\nu(\vec x\; \widetilde{\phi}_1(\,\cdot\,;\,\vec \beta),\vec m)
\end{equation}
The stray field was then computed with AD in forward mode
\begin{equation}
    \vec h_{ELM}(\vec x;\, \vec \beta) = -\nabla\phi_{ELM}(\vec x;\, \vec \beta).
\end{equation}
From this PINN solution and the exact solution, the error of the scalar potential and the stray field was computed. 

To compute the single layer potential, the surface of the sphere was parameterized by
\begin{equation}\label{eq:parametrization-sphere-surface}
    s(\varphi, \vartheta) = R\left( 
    \begin{array}{c}
        \sin(\varphi) \cos(\vartheta) \\
        \sin(\varphi) \sin(\vartheta) \\
        \cos(\varphi) \\ 
    \end{array} \right)\quad \varphi\in [0, \pi],\; \vartheta\in [0, 2\pi].
\end{equation}
The input space $[0, \pi]\times[0, 2\pi]$ was partitioned into $8\times8$ equal spaced tiles and mapped onto the surface of the sphere. For each sample point of the domain, the \textit{surface tensor} was pre-computed with a $4$\textsuperscript{th} order Gauss–Legendre quadrature and the \textit{charge tensor} was evaluated for the center points of the tiles in the parameter space $[0, \pi]\times[0, 2\pi]$. A second order Taylor expansion was used for the computation of the \textit{source tensor}. The training data was drawn from a Halton sequence \cite{halton1960efficiency}. In total, there were $4096$ collocation points which are uniformly drawn from $[0,1]^3$ and uniformly mapped onto the sphere with the parametrization
\begin{equation}\label{eq:parametrization-sphere}
        z(r, \varphi, \vartheta) = r\left( 
        \begin{array}{c}
            \cos(\vartheta) \sqrt{1 - \varphi ^ 2} \\
            \sin(\vartheta) \sqrt{1 - \varphi ^ 2} \\
            \varphi \\ 
        \end{array} \right),\;\;\;\varphi = 2 \vec x_1 - 1,\,\vartheta = 2 \pi \vec x_2,\,r = R \sqrt[3]{\vec x_3},\,\vec x \in [0,1]^3.
\end{equation}

\begin{figure}[ht]
    \centering
    \includegraphics[width=1\linewidth]{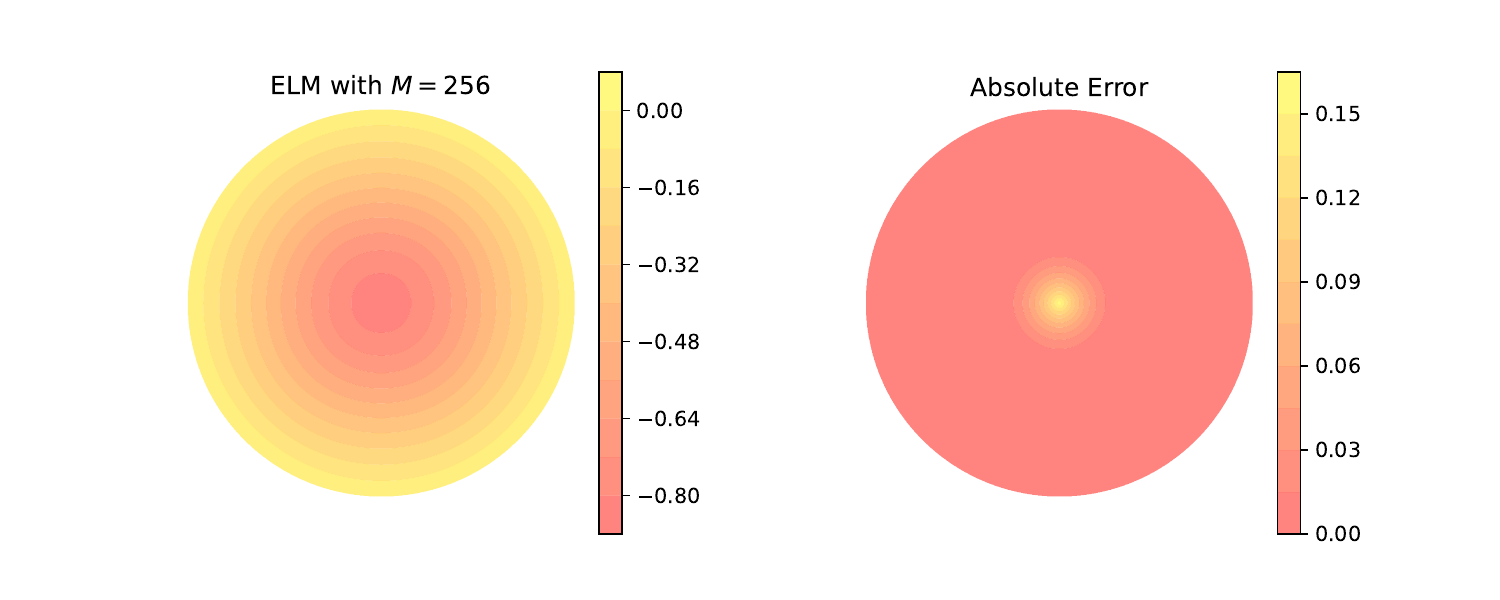}
    \caption[ELM solution of an outward magnetized sphere]{Solution in the $x-y$ plane of the scalar potential for an outward magnetized sphere, computed with an ELM with $256$ hidden nodes and \texttt{gelu} activation function, together with the absolute error of the solution.}
    \label{fig:sphere-elm256}
\end{figure}
\begin{figure}[ht]
    \centering
    \begin{subfigure}[b]{0.49\textwidth}
        \centering
        \includegraphics[width=\textwidth]{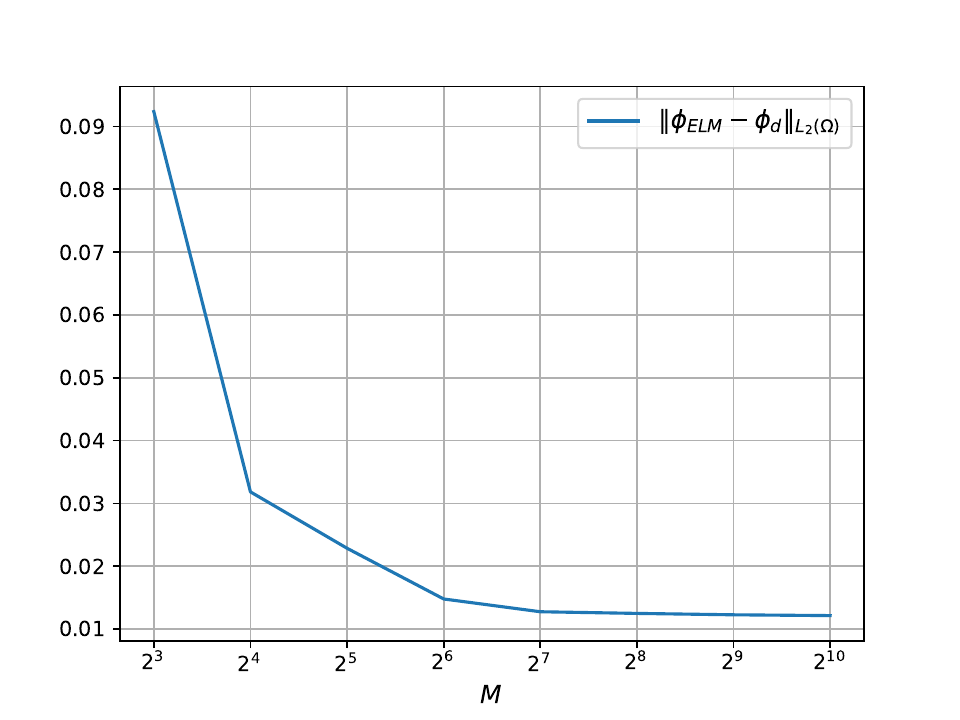}
    \end{subfigure}
    \hfill
    \begin{subfigure}[b]{0.49\textwidth}
        \centering
        \includegraphics[width=\textwidth]{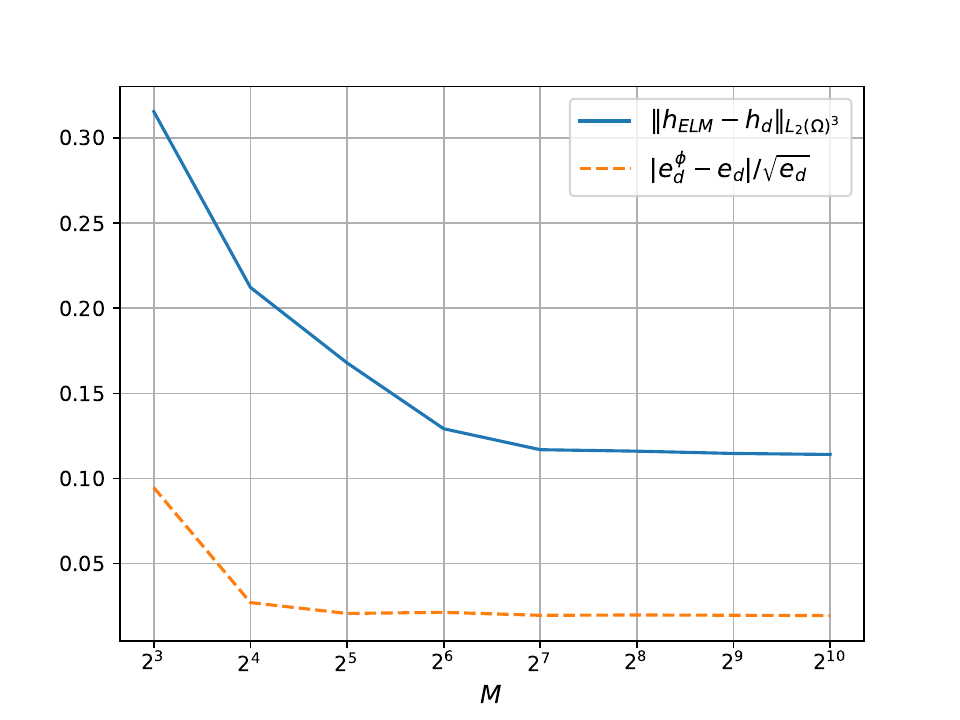}
    \end{subfigure}
    \caption[Error plots of an ELM solution of an outward magnetized sphere]{Error plots of the scalar potential for an outward magnetized sphere, computed with an ELM with \texttt{gelu} activation function for a varying number of hidden nodes.}
    \label{fig:sphere-error-elm}
\end{figure}
The ELM solution was computed for a variable hidden layer size $M \in \{8,\linebreak[1] 16,\linebreak[1] 32,\linebreak[1] 64,\linebreak[1] 128,\linebreak[1] 256,\linebreak[1] 512,\linebreak[1] 1024\}$.
The ELM weights were drawn with a Halton sequence from $[-2, 2]^4$, where the $4$\textsuperscript{th} dimension stems from the bias term of the ELM. A \texttt{gelu} activation function was used for all ELMs. Figure~\ref{fig:sphere-elm256} shows the solution of the scalar potential as well as the absolute error of an ELM with $256$ hidden nodes. It can be seen that the ELM has problems to approximate the solution at the center, since the solution is not smooth at the origin, but the model is. This is a general issue of the continuous PINN ansatz. The error plots for a varying number of hidden nodes are shown in Figure~\ref{fig:sphere-error-elm}. Also, the relative error of the energy is shown. The magnetostatic energy was computed as in \eqref{eq:demag-energies} with
\begin{equation}
    e_d^{\phi}(\vec\beta) = -\int_\Omega \vec m(\vec x) \cdot \vec h_{ELM}(\vec x;\,\vec\beta)\,\mathrm{d}\vec x,
\end{equation} 
on an independent validation data set of the same size as the training data set. It was drawn from a Halton Sequence as well.
It can be seen that the error estimates \eqref{eq:error-energy} hold true.

\subsection{Flower state}\label{sec:flower-state}
\begin{figure}[ht]
    \centering
    \begin{subfigure}[b]{0.49\textwidth}
        \centering
        \includegraphics[width=\textwidth]{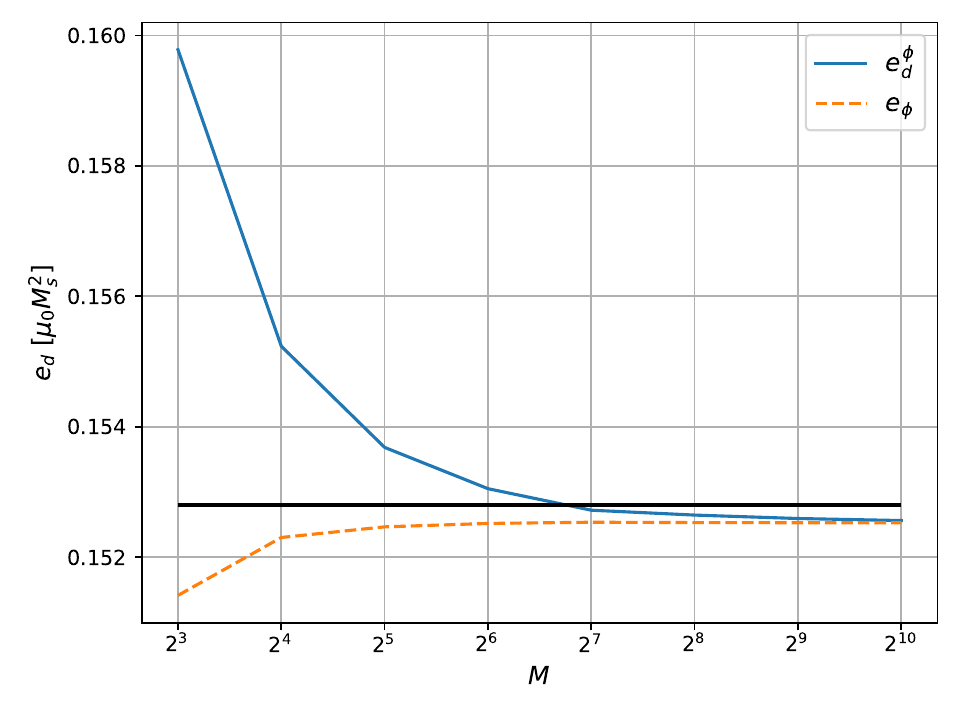}
    \end{subfigure}
    \hfill
    \begin{subfigure}[b]{0.49\textwidth}
        \centering
        \includegraphics[width=\textwidth]{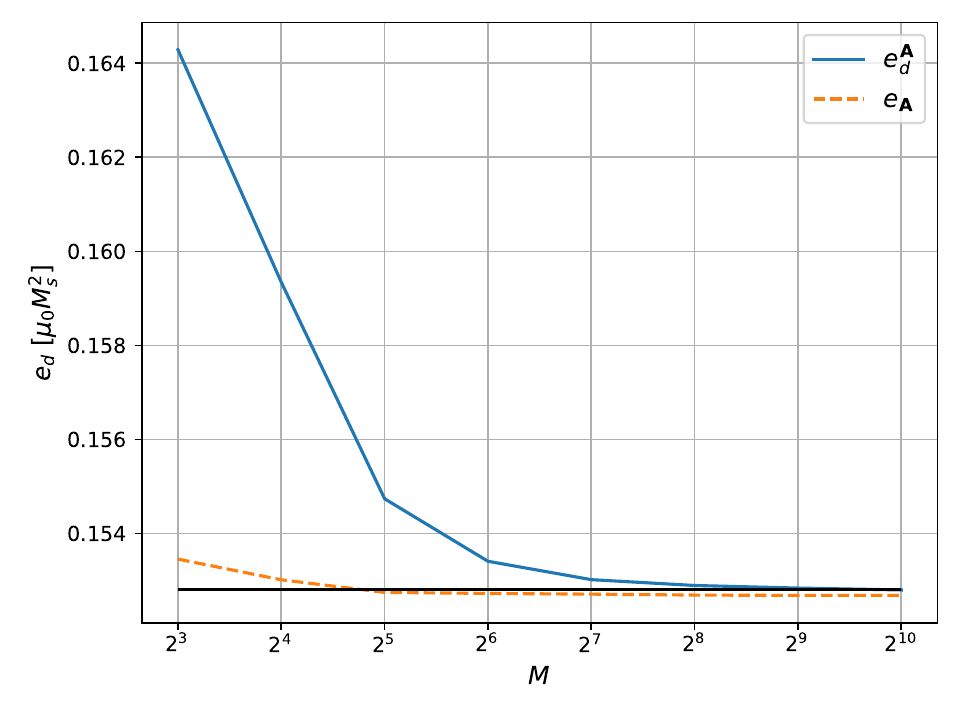}
    \end{subfigure}
    \caption[Magnetostatic energy and energy bounds for a flower state computed with ELMs]{Magnetostatic energy and energy bounds \eqref{eq:lowerbound-finite} and \eqref{eq:upperbound-finite} for a flower state  computed with an ELM with a variable number of hidden nodes, and \texttt{tanh} activation. The energy was computed with the scalar potential (left) and with the vector potential (right). The horizontal line marks the reference solution of $\num{0.1528}[\mu_0M_s^2]$ computed with the demagnetization tensor method with a discretization of $40\times 40\times 40$ cells \cite{abert2013numerical}.}
    \label{fig:cube-flower-state-elm}
\end{figure}

The second test for the computation of the scalar potential, as well as the vector potential, was a flower state in a unit cube $[-0.5, 0.5]^3$. The magnetization was given by the normalized version of
\begin{equation}
\vec m(\vec x) = \left(\vec x_1\vec x_3,\vec x_2\vec x_3 + (\frac{1}{2}\vec x_2 \vec x_3)^3,1\right)^T
\end{equation}
For the hard constraint model, the ADF 
\begin{equation}\label{eq:adf-cube}
    \ell(\vec x) = \hat\ell(\vec x_1) \sim_0 \hat\ell(\vec x_2) \sim_0 \hat\ell(\vec x_3),\quad \hat\ell(x) = (x + 0.5) \sim_0 (0.5 - x)
\end{equation}
was used, where $\sim_0$ is the $R_0$-equivalence operation \cite{sukumar2022exact}.
Again, a training set with $N=2^{12}$ was drawn from a Halton sequence. The ELM weights were also drawn from a Halton sequence from $[-2, 2]^4$ and a \texttt{tanh} activation function was used for the hidden layer. Further, an $\ell_2$-regularization term of $\mu=\num{e-3}$ was added to the ELM (see section \ref{sec:elm}). Each face of the surface is discretized with $6\times 6$ equal sized tiles, and a $10$\textsuperscript{th} order Gauss–Legendre quadrature rule was used to evaluate the single layer potential. The Taylor expansion was again of second order. By \textit{Monte Carlo} integration, the magnetostatic energy \eqref{eq:demag-energies} was computed with an independent validation set with $2^{13}$ samples. Also, the lower bounds \eqref{eq:lowerbound-finite} and upper bounds \eqref{eq:upperbound-finite} were computed. The double surface integrals were evaluated with \eqref{eq:phi1-double-surface-int} and \eqref{eq:A1-double-surface-int}. The results are shown in Figure~\ref{fig:cube-flower-state-elm}. The computed solution of the energy with respect to the scalar potential is denoted with $e_d^\phi$ and with respect to the vector potential with $e_d^{\vec A}$. The reference solution is $0.1528[\mu_0Ms^2]$, which was computed with the demagnetization tensor method with a discretization of $40\times 40\times 40$ cells \cite{abert2013numerical}. The upper bound is slightly lower than the reference solution. This error originates from numerical integration. 
It can be seen that the bounds and the respective solution converge to the same value, hence the bounds, especially the lower bound, might be used as a validation metric for the model. If the difference between the bound and the energy is too large, the model performance is likely to be bad.

As a further test, the solution was also approximated with a hard constraint PINN ansatz for the scalar potential as well as for the vector potential.
The network structure given in Table~\ref{tab:pinn-flower-state}.
\begin{table}[ht]
    \centering
    \begin{tabular}{l|c|c|c}
        Layer & Activation & Input shape & Output shape \\
        \hline
        Dense & \texttt{gelu} & 3 & 16 \\
        Dense & \texttt{gelu} & 16 & 16 \\
        Dense & \texttt{identity} & 16 & $t$ \\
    \end{tabular}
    \caption[PINN architecture for the flower state model]{PINN architecture for the flower state model. The output dimension was $t=1$ for the scalar potential and $t=3$ for the vector potential.}
    \label{tab:pinn-flower-state}
\end{table}
\begin{figure}[ht]
    \centering
    \begin{subfigure}[b]{0.49\textwidth}
        \centering
        \includegraphics[width=\textwidth]{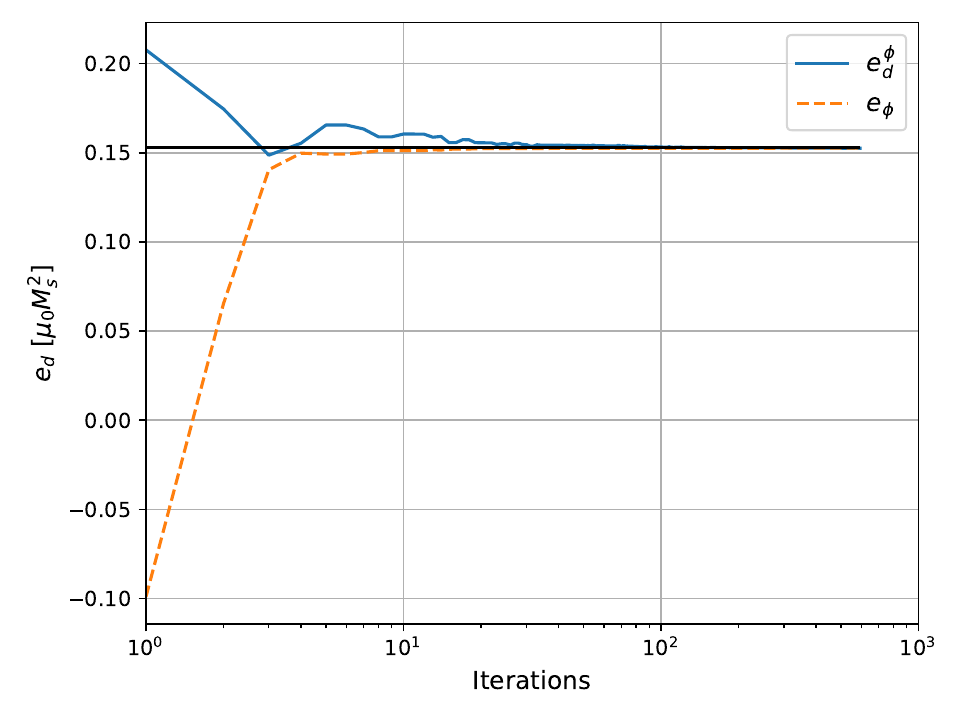}
    \end{subfigure}
    \hfill
    \begin{subfigure}[b]{0.49\textwidth}
        \centering
        \includegraphics[width=\textwidth]{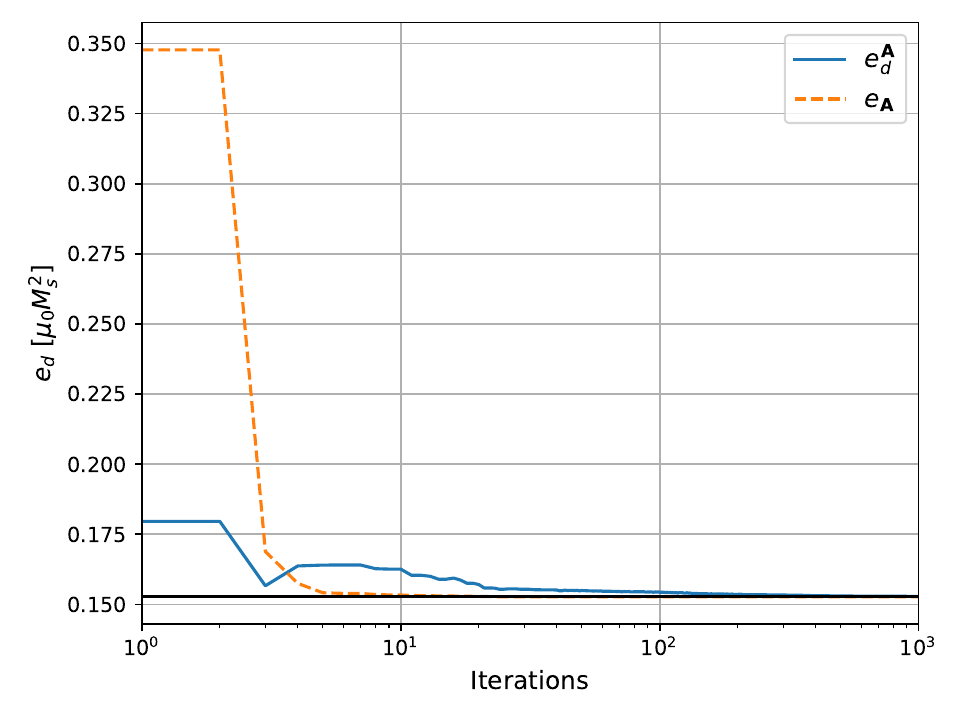}
    \end{subfigure}
    \caption[Magnetostatic energy and energy bounds for a flower state computed with PINNs]{Magnetostatic energy and energy bounds \eqref{eq:lowerbound-finite} and \eqref{eq:upperbound-finite}  for a flower state computed with a hard constraint PINN. A trust region optimizer was used, and the curves display the energies and energy bounds during the training process. The network architecture is given in Table~\ref{tab:pinn-flower-state}. The energy was computed with the scalar potential (left) and with the vector potential (right). The horizontal line marks the reference solution of $\num{0.1528}[\mu_0M_s^2]$ computed with the demagnetization tensor method with a discretization of $40\times 40\times 40$ cells \cite{abert2013numerical}.}
    \label{fig:cube-flower-state-pinn}
\end{figure}
Figure~\ref{fig:cube-flower-state-pinn} shows the energies and energy bounds during the training process. A trust region optimizer with the Steihaug Conjugate Gradient (Steihaug-CG) as a subproblem solver was used for training.

Table~\ref{tab:flower-state-energies} shows the computed self energies and energy bounds of the final model of the PINNs and the largest ELMs with $M=1024$, i.e., the rightmost values of Figure~\ref{fig:cube-flower-state-elm} and Figure~\ref{fig:cube-flower-state-pinn}. It can be seen that the ELM and PINN both compute the solution very accurately, but training time for the ELM was only a fraction of a second, while for the PINN it takes minutes. However, this strongly depends on the optimizer and the required accuracy.
\begin{table}[ht]
    \centering
    \begin{tabular}{l | c | c | c | c}
        & $e_d^\phi$ & $e_d^{\vec A}$ & $e_\phi$ & $e_{\vec A}$  \\
        \hline
        ELM $M=1024$ & \num{0.15256} & \num{0.15290} & \num{0.15253} & \num{0.15268} \\
        PINN         & \num{0.15249} & \num{0.15287} & \num{0.15256} & \num{0.15273}
    \end{tabular}
    \caption[Computed self energies and energy bounds of a flower state in a cube for a hard constraint ELM and a hard constraint PINN model]{Computed self energies and energy bounds of a flower state in a cube for a hard constraint ELM and a hard constraint PINN model. The reference solution is $\num{0.1528}[\mu_0M_s^2]$, computed with a demagnetization tensor method with a discretization of $40\times 40\times 40$ cells \cite{abert2013numerical}.}
    \label{tab:flower-state-energies}
\end{table}

\subsection{Vortex state}\label{sec:vortex-state}
\begin{figure}
    \centering
    \includegraphics[width=1.0\linewidth]{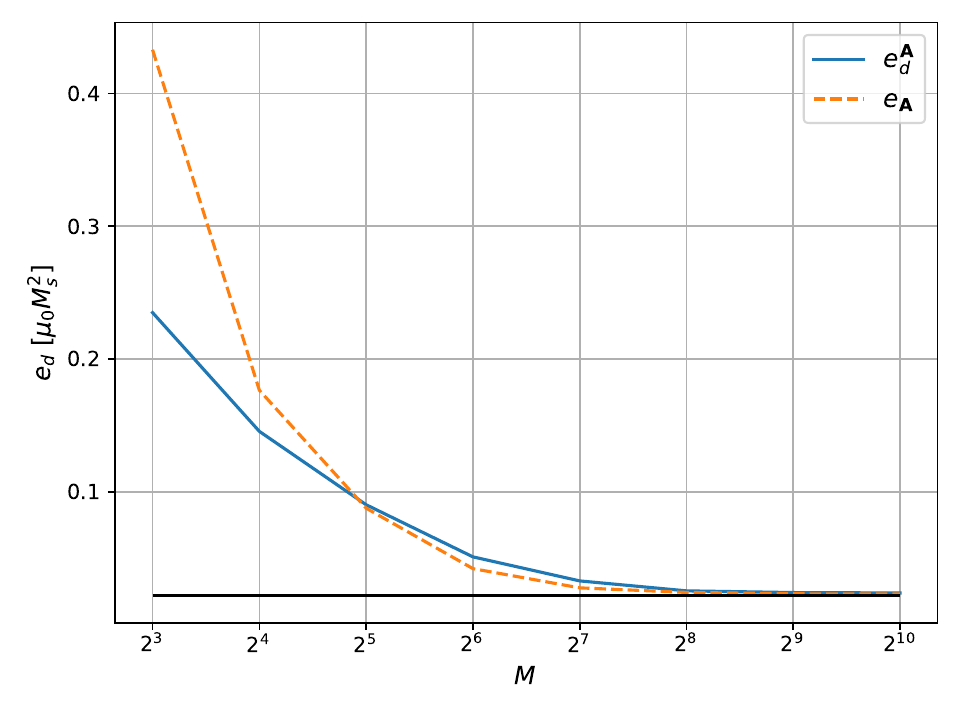}
    \caption[Magnetostatic energy and the upper energy bound for a vortex state computed with ELMs and vector potential]{Magnetostatic energy and the upper energy bound \eqref{eq:upperbound-finite} computed with an ELM via vector potential with a variable number of hidden nodes, and \texttt{RBF} activation, where the kernel parameter $\gamma=10$ was used. The horizontal line marks the reference solution of $\num{0.0219}[\mu_0M_s^2]$ computed with the demagnetization tensor method with a discretization of $40\times 40\times 40$ cells \cite{abert2013numerical}. The energy for the ELM with $2^{10}$ hidden nodes was $e_d^{\vec A}=0.0237[\mu_0M_s^2]$ and the upper bound $e_{\vec A}=0.0229[\mu_0M_s^2]$.}
    \label{fig:cube-vortex-state-elm}
\end{figure}

For this test, a vortex state within a cubic domain is solved via vector potential. The scalar potential is not considered since it has a trivial solution $\phi_1=0$. The magnetization is given by the normalized version of 
\begin{equation}
    \vec m_v(\vec x) = \left(- \frac{\vec x_3}{r} \sqrt{1-\exp\left(-4\frac{r^2}{r_c^2}\right)}, \exp\left(-2\frac{r^2}{r_c^2}\right), \frac{\vec x_1}{r} \sqrt{1-\exp\left(-4\frac{r^2}{r_c^2}\right)}\right)^T,
\end{equation}
with $r = \sqrt{\vec x_1^2+\vec x_3^2}$ and $r_c=0.14$. The domain, ADF, evaluation of the single layer potential and the sample set is the same as for the flower state. A hard constrained ELM with \texttt{rbf} activation $\sigma_i(\vec x) = \exp(-\gamma \|\vec x - \vec w_i\|^2)$ with $\vec w_i \in  \Omega$ and $\gamma=10$ was solved for a variable number of hidden nodes. The reference solution is $0.0219[\mu_0Ms^2]$, which was computed with the demagnetization tensor method with a discretization of $40\times 40\times 40$ cells \cite{abert2013numerical}. Figure~\ref{fig:cube-vortex-state-elm} shows the result of this computation. The energy for the ELM with $2^{10}$ hidden nodes was $e_d^{\vec A}=0.0237[\mu_0M_s^2]$ and the upper bound $e_{\vec A}=0.0229[\mu_0M_s^2]$. This problem is quite difficult to solve, and we were not able to solve this problem with other ELM activations. This indicates that \texttt{rbf} activations are very powerful function approximators, however, for easy functions other choices also work incredible well. Further results in this section were still achieved with \texttt{tanh} activations.

\subsection{Demagnetization of a hard magnetic sphere}\label{sec:sphere-energy-min}
The second test involved the computation of the demagnetization process of a hard magnetic $Nd_2Fe_{14}B$ sphere with radius $R=\qty{10}{nm}$. The main purpose was to test the capabilities of the hard constrained neural network model to capture the demagnetization and the switching of the magnetization. The sphere had uniaxial anisotropy in $z$-direction, with the anisotropy constant $K_u = \qty{4.3e6}{J/m}$. Saturation polarization for the hard magnetic material was  $J_s=\qty{1.61}{T}$ and the exchange stiffness for $Nd_2Fe_{14}B$ was selected to be $A_{ex}=\qty{7.3e-12}{J/m}$. Therefore, the reduced uniaxial anisotropy constant was $Q=K_u / K_m \approx 2.085$. The demagnetization was computed with an applied field along the anisotropy axis $(0,0,1)^T$ with the analytic value of the switching field given by $h_{sw}=2Q$ and with the applied field in a $\ang{45}$ angle to the anisotropy axis with the analytic value of the switching field  $h_{sw}=Q$ \cite{kronmuller2007general}. The micromagnetic self energy was computed with the vector potential, which is modeled with a PINN as given in Table~\ref{tab:pinn-vec-pot-sphere}. The ADF was given by $\ell(\vec x) = (R^2 - \|\vec x\| ^ 2) / 2R)$. A hard constrained ansatz, as in \eqref{eq:mag_model}, was used to model the magnetization and the network architecture for $\mathcal{N}_{\vec\theta}$ was the same as in Table~\ref{tab:pinn-vec-pot-sphere}. The initial magnetization $\vec m_0$ was uniform along the anisotropy axis.
\begin{table}[ht]
    \centering
    \begin{tabular}{l|c|c|c}
        Layer & Activation & Input shape & Output shape \\
        \hline
        Dense & \texttt{gelu} & 3 & 12 \\
        Dense & \texttt{gelu} & 12 & 12 \\
        Dense & \texttt{identity} & 12 & 3 \\
    \end{tabular}
    \caption[Neural network model for the vector potential of a hard magnetic sphere]{Neural network model for the vector potential of a hard magnetic $Nd_2Fe_{14}B$ particle.}
    \label{tab:pinn-vec-pot-sphere}
\end{table}
\begin{figure}
    \centering
    \includegraphics[width=\linewidth]{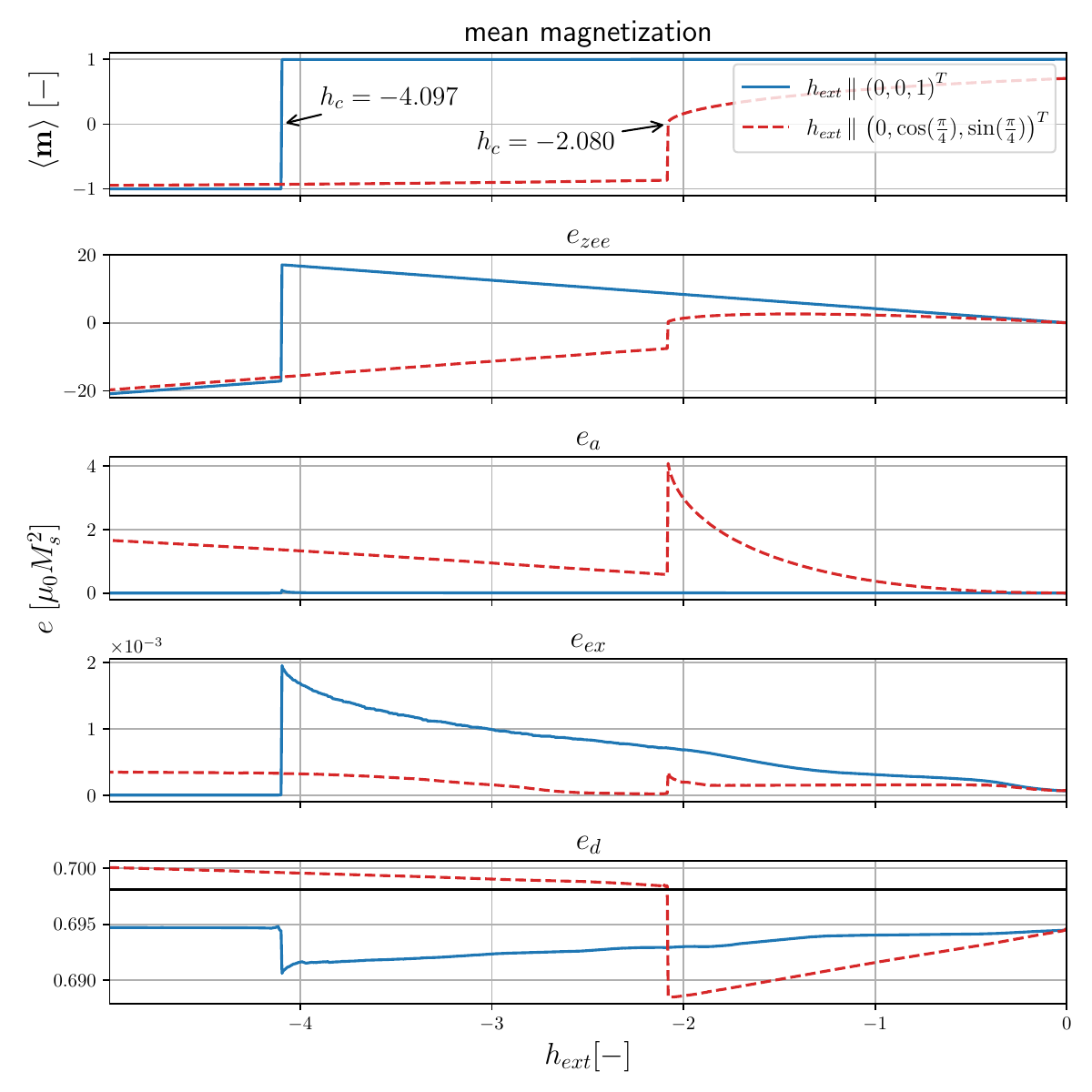}
    \caption[Demagnetization process of a hard magnetic sphere]{Demagnetization process of a hard magnetic sphere with the applied field along $(0,0,1)^T$ and $(0, \cos(\pi/4), \sin(\pi/4))^T$. The mean magnetization was computed by projection onto the axis of the applied field. The horizontal line in the last plot highlights the magnetostatic energy of a uniformly magnetized sphere.}
    \label{fig:demag-sphere-vec-pot}
\end{figure}
The computational domain was scaled to the unit sphere. The scaled exchange constant was given by  $\widetilde{A}_{ex}'=\widetilde{A}_{ex} / R^2$. Energy minimization was performed with Algorithm~\ref{alg:phi1-ed-min} but with the vector potential and $\check e$. For the evaluation of the single layer potential, the parametrization \eqref{eq:parametrization-sphere-surface} was used, with the input space partitioned into $6\times6$ equal spaced tiles. Each tile was integrated with a $5$\textsuperscript{th} order Gauss–Legendre quadrature rule. In total, there were $4096$ collocation points, drawn from a Halton sequence \cite{halton1960efficiency}, and are mapped to the domain with the parametrization \eqref{eq:parametrization-sphere}. A Trust Region optimizer, with Steihaug-CG method as a sub-problem solver, was used for minimization and the parameters for Algorithm~\ref{alg:phi1-ed-min} were: $\epsilon_{\vec A_1}=\num{e-2}$, $\epsilon_{\vec m}=\num{e-2}$, $N_{\vec m} = 10$ and $N_{\vec A_1} = 20$. Also, a damping factor of $\num{e-3}$ was applied to regularize the Hessian matrix.
In the beginning of the demagnetization process, the particle was uniformly magnetized in the direction of the anisotropy axis and the applied field was set to zero. For each step, the applied field was reduced by $5e-3$. 

Figure~\ref{fig:demag-sphere-vec-pot} shows the demagnetization and the corresponding energy terms for both trials. The self-energy of a uniformly magnetized sphere is $4 \pi / 18\,\left[\mu_0M_s^2\right]$. It can be seen from the figure that the magnetostatic energy is very close to this value. From this and the exchange energy, we can conclude that the model performs quite well in approximating the magnetization reversal, which is a homogeneous rotation of the uniform magnetization. The computation was performed with single precision on a \textit{Geforce RTX 2080 Ti}.

\subsection{Demagnetization of a hard magnetic cube}\label{sec:cube-energy-min}
We compute the demagnetization process of a hard magnetic $Nd_2Fe_{14}B$ cube with an edge length of $L=\qty{70}{nm}$. The material parameters and anisotropy axis were the same as for the sphere, and the scaled exchange constant was given by  $\widetilde{A}_{ex}'=\widetilde{A}_{ex} / L^2$. With energy density $K_m = \mu_0M_s^2$ J/m, the exchange length is $\ell_{ex}=\sqrt{2A_{ex}/K_m} \approx \qty{2.6}{nm}$, the material parameter $Q=K_u / K_m \approx 2.08$ which gives a wall width parameter of $\sqrt{A_{ex}/K_u}\approx \qty{1.3}{nm}$. The external field was applied along the vector $(1,0,10)^T$ and was nearly parallel to the anisotropy axis. 

The neural network architecture of $\mathcal{N}_{\vec\theta}$ to model the magnetization is given in Table~\ref{tab:nn-model-demag-cube}. The initial magnetization was uniform along the anisotropy axis.
\begin{table}
    \centering
    \begin{tabular}{l|c|c|c}
        Layer & Activation & Input shape & Output shape \\
        \hline
        Dense & \texttt{gelu} & 3 & 16 \\
        Dense & \texttt{gelu} & 16 & 16 \\
        Dense & \texttt{identity} & 16 & 3 \\
    \end{tabular}
    \caption[Neural network model for the computation of a demagnetization process for a hard magnetic cube]{Neural network model for the computation of a demagnetization process for a $Nd_2Fe_{14}B$ cube.}
    \label{tab:nn-model-demag-cube}
\end{table}
The training set contained $2^{12}$ samples from a Halton sequence. For a decreasing external field $h_{ext} = \|\vec h_{ext}\|\in[-3.5, 1]$, an equilibrium state was computed for each step using a trust region method. The constrained quadratic trust region problem was solved with Steihaug-CG method and a damping factor of $\num{e-3}$ was used for Hessian regularization. At each step $h_{ext}$ was reduced by $\Delta h_{ext}=\num{5e-3}$. The magnetostatic energy was modeled via scalar and vector potential. 

\subsubsection{ELM ansatz}\label{sec:cube-elm}
\begin{figure}
    \centering
    \includegraphics[width=\linewidth]{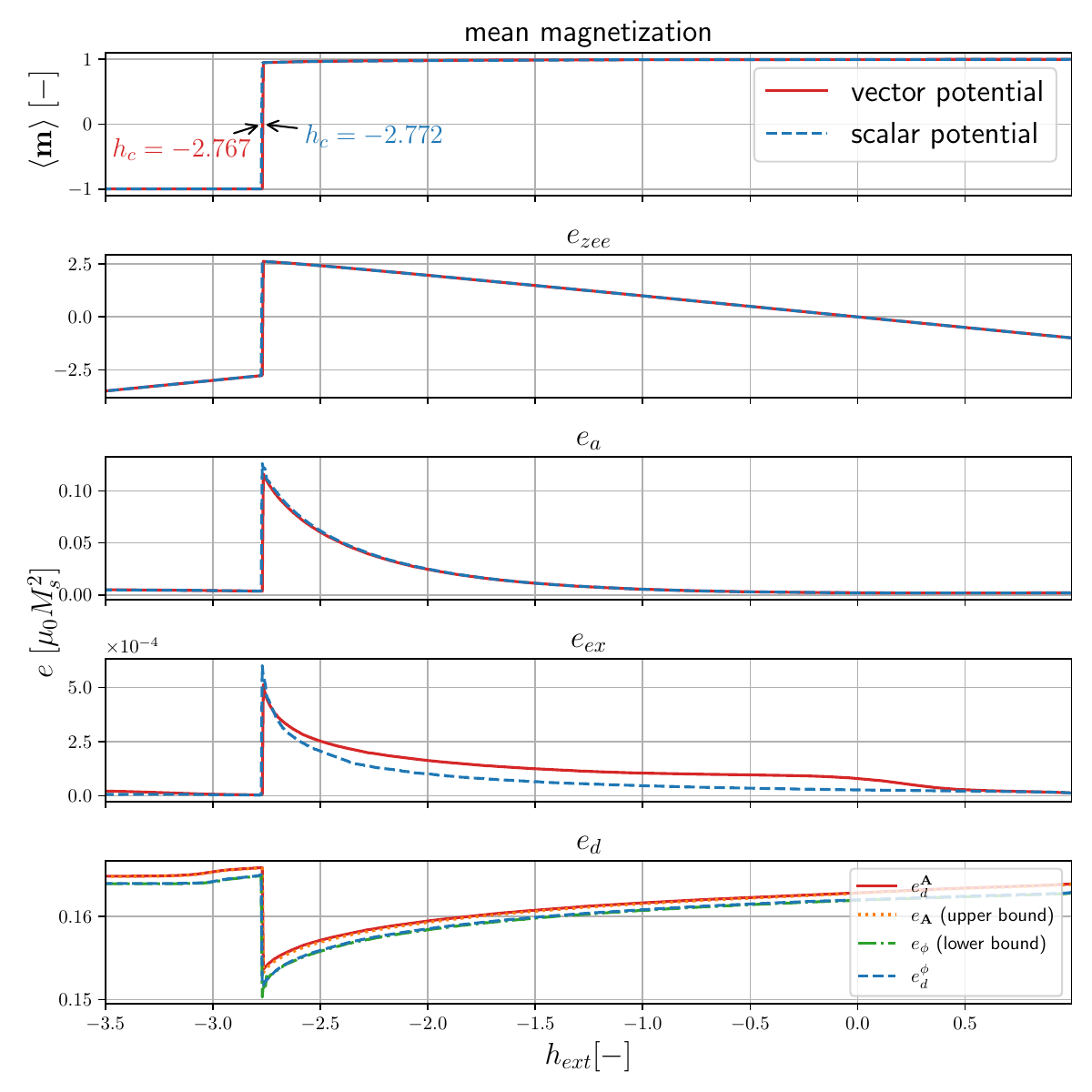}
    \caption[Demagnetization process of a hard magnetic cube with ELMs for computation of scalar and vector potential]{Demagnetization curve of a 70nm $Nd_2Fe_{14}B$ cube. The magnetization was projected onto the axis $(1,0,10)^T$ of the applied field. The computation was performed with a hard constrained ELM model for scalar and vector potential. }
    \label{fig:demag-cube-elm}
\end{figure}

In a first trial, an ELM with $512$ hidden nodes, \texttt{tanh} activation and the input weights drawn from $[-2, 2]^4$, was applied to model $\phi_1$ and $\vec A_1$. The solution operator was precomputed with a small Ridge regularization of $\num{e-3}$ and was the same for scalar and vector potential. The ADF and the evaluation strategy for the single layer potential was the same as in Section~\ref{sec:flower-state}. 
Algorithm~\ref{alg:energy-minimization2} with $\epsilon_{\vec m} = \num{e-3}$ was used for energy minimization. Figure~\ref{fig:demag-cube-elm} shows the demagnetization process for the ELM ansatz for vector and scalar potential. Also, lower \eqref{eq:lowerbound-finite} and upper bounds \eqref{eq:upperbound-finite} were computed, and it can be seen that energies and bounds were almost identical, which indicates that the model for computation of the self-energy was sufficient. The switching field was found at $h_c=-2.772$ for the scalar potential version and at $h_c=-2.767$ for the vector potential, which is the same as given in the reference \cite{exl2014tensor}. The scalar potential missed the switching by a single step, which is still a very good result. One can see that the exchange energy of the vector potential simulation is higher than the one of scalar potential simulation, which is an artifact of the optimization process.

In our trials, we found that the improper evaluation of the single layer potential can have a negative effect on the computation and stability of the minimization, leading to early switching. Some validation metric of an accurate evaluation would therefore be of interest.

During the switching process, the energy landscape has a saddle point. Therefore, we assume that the quadratic form of the trust region method can capture this behavior very well during optimization \cite{schaffer2024constraint}. 

\subsubsection{L-BFGS Optimizer}
\begin{figure}[ht]
    \centering
    \includegraphics[width=\linewidth]{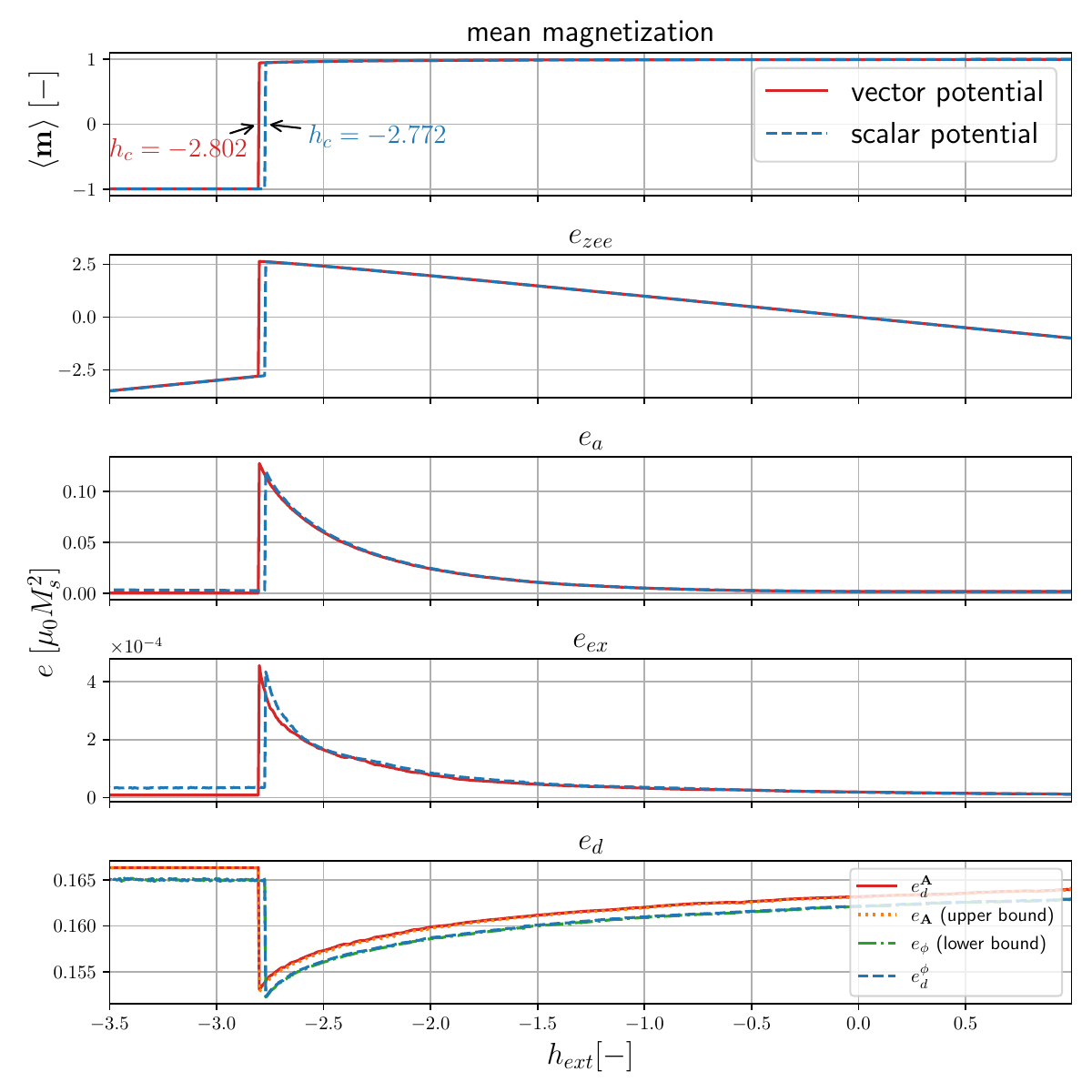}
    \caption[Demagnetization process of a hard magnetic cube with ELMs for computation of scalar and vector potential and L-BFGS optimizer]{Demagnetization curve of a 70nm $Nd_2Fe_{14}B$ cube computed with an L-BFGS optimizer. The magnetization was projected onto the axis $(1,0,10)^T$ of the applied field. The computation was performed with a hard constrained ELM model for scalar and vector potential. }
    \label{fig:demag-cube-elm-lbfgs}
\end{figure}

Although the trust region framework works reliably for energy minimization, the high computational cost to solve the subproblem is problematic. Also, if the parameters are already close to the optimum, the method struggles to find a good optimum in the flat energy landscape, leading to slow convergence. Therefore, line search methods, especially L-BFGS \cite{nocedal1999numerical}, seem to be more suitable for this task. The same test was performed with an L-BFGS optimizer and the same ELM ansatz for scalar and vector potential. While for the trust region method, the scalar and vector potential was evaluated every iteration, $N_{\vec m}=20$ L-BFGS iterations are performed before re-evaluating the potential. Figure~\ref{fig:demag-cube-elm-lbfgs} shows the result of the computation. The switching occurred at $h_c = -2.772$ for the scalar potential ansatz and at $h_c = -2.802$ for the vector potential, while the computation took only about $\qty{90}{s}$. This shows that also a positive definite Hessian approximation with line search can capture the switching. However, occasionally, the switching might not be completed after a single optimization step, which can cause the optimizer to stall since the history does not match. The information saved in the history of the previous iterates are a very poor approximation of the Hessian. This requires a restart of the optimizer, which is also necessary after the switching is completed.
The trust region method seems to be the better choice to compute the switching field accurately, whereas L-BFGS outperforms the trust region method in terms of computation time. This suggests that both optimizers could be applied together. This could be done by switching to the trust region scheme close to the switching point. Another possibility would be the usage of a sampled version of L-BFGS as in \cite{berahas2019quasi} which should eliminate this issue. It might also be possible to intelligently sample the respective objective function during line search.

\subsubsection{PINN ansatz}\label{sec:cube-pinn}
\begin{figure}[ht]
    \centering
    \includegraphics[width=\linewidth]{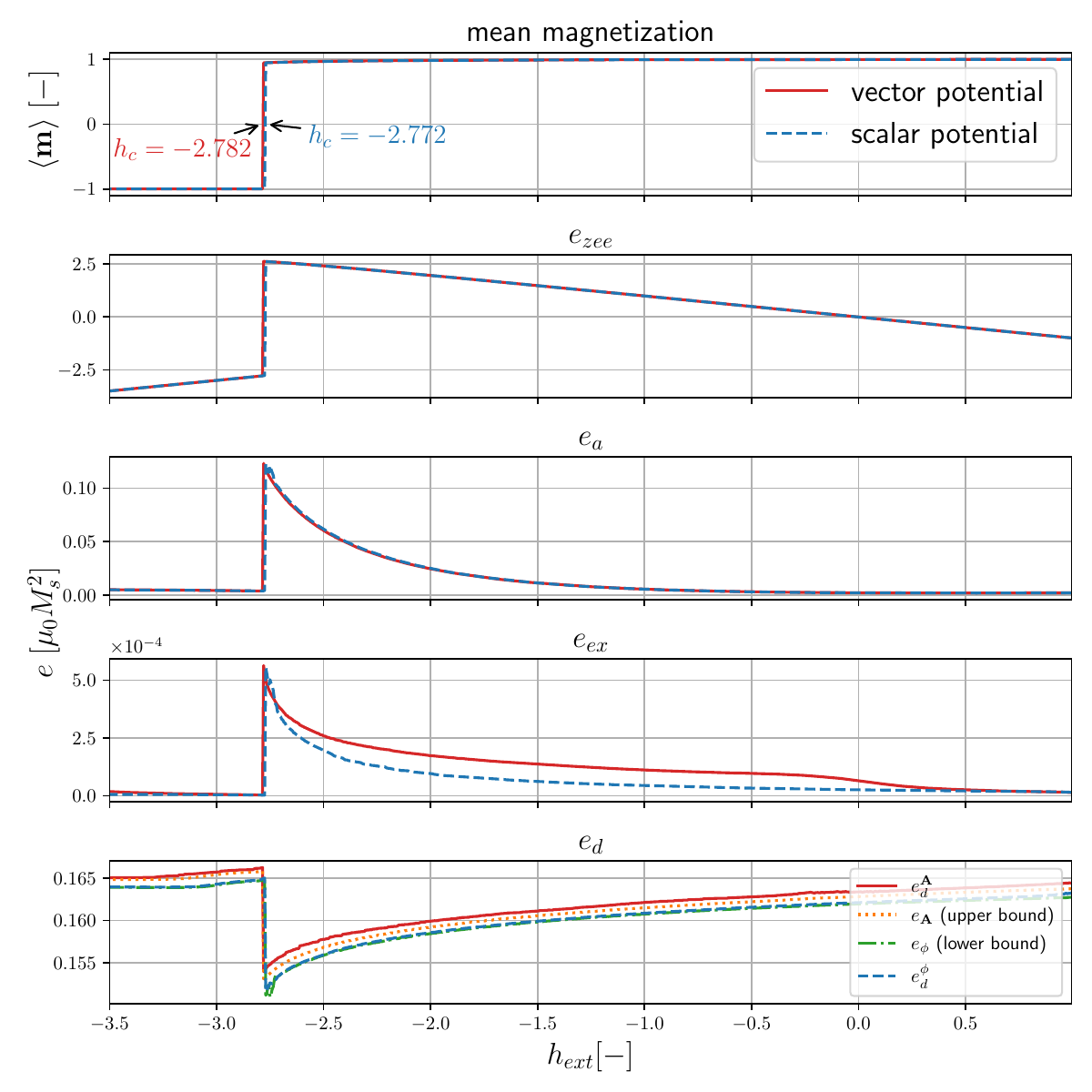}
    \caption[Demagnetization process of a hard magnetic cube with PINN ansatz for computation of scalar and vector potential]{Demagnetization curve of a 70nm $Nd_2Fe_{14}B$ cube. The magnetization was projected onto the axis $(1,0,10)^T$ of the applied field. The computation was performed with a hard constrained PINN model for scalar and vector potential.}
    \label{fig:demag-cube}
\end{figure}

The same test was performed with a hard constrained PINN ansatz for scalar and vector potential. The model was the same as for the magnetization in Table~\ref{tab:nn-model-demag-cube} (the scalar potential has scalar output). Minimization was performed with Algorithm~\ref{alg:phi1-ed-min} for the scalar potential and the respective version for the vector potential and trust region optimization. The parameters were $N_{\vec m} = 1$, $N_{\phi_1} = N_{\vec A_1} = 20$, $\epsilon_{\vec m} = \num{e-3}$ and $\epsilon_{\phi_1} = \epsilon_{\vec A_1} = \num{5e-2}$. Figure~\ref{fig:demag-cube} shows the result for this version. The switching field is the same for the scalar potential as in the last trial, but is slightly off for the vector potential, $h_c = -2.782$. The plot for the self energy shows that the upper bound is slightly lower than the computed energy with the vector potential. This could explain why the result was worse. The vector potential might not be accurate enough and one would need to decrease $\epsilon_{\vec A_1}$ or use another network architecture to improve the result. 

\section{Conclusion}
In this work, we have explored the application of machine learning techniques, such as Physics-Informed Neural Networks (PINNs) and Extreme Learning Machines (ELMs), to address the challenges of magnetostatics, particularly in the context of Gibbs free energy minimization.
We present an alternating optimization method for the joint optimization of Brown's energy bounds and the Gibbs free energy. Our method enforces the unit norm constraint via the Cayley transform on the model, removing the need for an expensive penalty framework. Additionally, building on the work of Garcia-Cervera and Roma, we can reformulate the energy bounds on the finite domain, which allows for more efficient computation of magnetostatic fields without having to account for external space.

While machine learning methods like PINNs and ELMs provide promising alternatives to traditional methods, several challenges remain. This includes hyperparameter tuning and efficient optimization of PINN models. We find, that the enforcement of essential boundary conditions via $R$-functions can result in highly accurate models which can compete with traditional numerical methods. Optimization of hard constraint PINNs can be challenging. Hard constraint ELMs, on the other hand, offer an easy and straightforward solution process.
In addition, we introduce the kernel and neural operator perspective for these cases, opening up novel algorithmic directions and applications. 

We tested these methods for the computation of the magnetostatic fields and the demagnetization curve for simple domains.
Future research focuses on refining these methods and apply them to more complex geometries, such as grain structure models. Further, DeepONets are of particular interest for the fast prediction of magnetostatic fields.

\section*{Acknowledgements}
\noindent Financial support by the Austrian Science Fund (FWF) via project ”Data-driven Reduced Order Approaches for Micromagnetism (Data-ROAM)” (Grant-DOI: 10.55776/PAT7615923), project ”Design of Nanocomposite Magnets by
Machine Learning (DeNaMML)” (Grant-DOI: 10.55776/P35413) is gratefully acknowledged. The authors acknowledge the University of Vienna research platform MMM Mathematics - Magnetism - Materials. The computations were partly achieved by using the Vienna Scientific Cluster (VSC) via the funded projects No. 71140 and 71952.
This research was funded in whole or in part by the Austrian Science Fund (FWF) [10.55776/PAT7615923, 10.55776/P35413]. For the purpose of Open Access, the author has applied a CC BY public copyright license to any Author Accepted Manuscript (AAM) version arising from this submission. 

\end{document}